\documentclass[%
final,
reprint,
superscriptaddress,
amsmath,amssymb,
aps,
prl,
floatfix,
nofootinbib,
]{revtex4-1}

\usepackage{color}

\usepackage[normalem]{ulem}
\usepackage{mystyle}
\usepackage{macros}

\begin{document}

\title{Lighting the Dark: The Evolution of the Post-Inflationary Universe}
\author{Nathan Musoke}
\email{n.musoke@auckland.ac.nz}
\author{Shaun Hotchkiss}
\email{s.hotchkiss@auckland.ac.nz}
\author{Richard Easther}
\email{r.easther@auckland.ac.nz}
\affiliation{Department of Physics, The University of Auckland, Private Bag 92019, Auckland, New Zealand}

\date{September 26, 2019}

\begin{abstract}
    In  simple inflationary cosmological scenarios the near-exponential growth can be followed by a long period in which the Universe is dominated by the oscillating inflaton condensate. The condensate is initially almost homogeneous, but  perturbations  grow gravitationally, eventually fragmenting the condensate if it is not disrupted more quickly by resonance or prompt reheating.  We show that the gravitational fragmentation of the condensate is well-described by the \SP{} equations and use numerical solutions to show that large overdensities form quickly after the  onset of nonlinearity. This is the first  exploration of this phase of nonlinear dynamics in the very early universe, which can affect the detailed form of the inflationary power spectrum and the dark matter fraction when the dark sector is directly coupled to the inflaton.
 \end{abstract}
\maketitle

\section{Introduction}
\label{sec:introduction}

In simple inflationary scenarios the universe grows at least $10^{60}$ times larger between the Planck scale and the present day~\cite{Guth:1980zm, Linde:1981mu, Albrecht:1982wi,Baumann:2009ds,Tanabashi:2018oca}.
Roughly speaking, the  inflationary phase  accounts for 30 of these 60 factors of 10, smoothing away  pre-inflationary remnants, and  laying down the perturbations that seed galaxy formation and the  anisotropies in the microwave sky.
In the  trillionth of a second after inflation, the universe grows by another 15 factors of 10, at which point, typical  interactions take place at Large Hadron Collider energies. The remaining growth occurs during the 13.8 billion years elapsing between that point and the present day.  Roughly half (logarithmically) of the post-inflationary growth of the universe occurs in an eye-blink and at energies beyond the reach of current experiments. This epoch is the primordial dark age~\cite{Boyle:2005se}: its unknown dynamics are critical to understanding both ultra-high-energy particle physics and the infant universe.

The universe must thermalise before neutrino freeze-out and can do so via several  mechanisms.
The inflaton condensate can fragment into its own quanta via self-resonance and particles coupled  to the inflaton can be resonantly produced off the condensate~\cite{Kofman:1994rk,Shtanov:1994ce,Kofman:1997yn}, leading to prompt thermalisation~\cite{Lozanov:2016hid} or a possible oscillon-dominated epoch~\cite{Amin:2010xe,Amin:2010dc,Amin:2011hj,Lozanov:2017hjm,Lozanov:2019ylm}. The full dynamics of this phase must typically be simulated via three dimensional Klein-Gordon solvers in a rigid, expanding spacetime~\cite{Felder:2000hq,Frolov:2008hy,Easther:2010qz}. Without resonance, particles are generated by slower, perturbative processes~\cite{Abbott:1982hn,Dolgov:1982th, Albrecht:1982mp}.
This perturbative mechanism takes places during a lengthy period of expansion in which the Universe is dominated by a coherent, nearly homogeneous condensate~\cite{Turner:1983he}, which eventually fragments via the gravitational growth of perturbations~\cite{Jedamzik:2010dq,Easther:2010mr}.
However, this process has not been studied beyond the breakdown of perturbation theory. Most treatments of resonance ignore local gravitational effects while fully relativistic solvers~\cite{Clough:2015sqa,East:2015ggf,Clough:2017ixw}  need to resolve the ``fast'' dynamics of the oscillating field, which is prohibitively expensive.

We show  that the  growing perturbations in the inflaton condensate are well described by the Schr\"{o}dinger-Poisson equations~\cite{Ruffini:1969qy,Spiegel:1980ykb,Seidel:1990jh,Widrow:1993qq,Woo:2008nn} and can be simulated with  tools~\cite{Woo:2008nn,Schive:2014dra,Schwabe:2016rze,Mocz:2017wlg,Edwards:2018ccc,Li:2018kyk,Zhang:2018ghp} similar to those used with ultralight dark matter  [ULDM]~\cite{Hu:2000ke, Marsh:2013ywa, Marsh:2015xka, Hui:2016ltb} or self-interacting axions \cite{Amin:2019ums}.
We  simulate the  fragmentation of a coherently oscillating inflaton condensate into localised, gravitationally confined overdensities,  a key step toward understanding of the primordial dark age in this  class of models.
This is key to fixing the detailed inflationary perturbation spectra~\cite{Dodelson:2003vq, Liddle:2003as, Adshead:2010mc, Munoz:2014eqa} and for making quantitative predictions in models where the dark matter is  coupled to the inflaton (e.g.~\cite{Chung:1998ua,Easther:2013nga,Fan:2013faa,Tenkanen:2016jic,Tenkanen:2019cik}) or consists of remnants of the inflaton itself~\cite{Liddle:2006qz,Tenkanen:2016twd,Hooper:2018buz,Almeida:2018oid}.

\section{Scenario}
\label{sec:scenario}
The inflaton, $\phi$, obeys the equation of motion
        \begin{equation}
            \label{eq:KG}
            \nabla_{\mu} \nabla^{\mu} \phi - V'( \phi) = 0 \,
        \end{equation}
and the metric obeys the Einstein field equations. In the homogeneous limit,  these reduce to
\begin{equation}
\ddot{\phi} + 3 H \dot{\phi} + \frac{d V}{d \phi} = 0\, ,\quad H^2 = \frac{1}{3 \Mpl^2} \left(\frac{{\dot \phi}^2}{2}
 + V(\phi)\right) \, ,
 \end{equation} 
where $\Mpl = \sqrt{1/8 \pi G}$ is the reduced Planck mass, $H = \dot{a}/a$, $a$ is the usual  scale factor, and $V$ is the potential.
Current  constraints require $V(\phi)$ to be sub-quadratic at \emph{large} field values~\cite{Akrami:2018odb}.
Non-quadratic potentials can induce self-resonance, fragmenting the inflaton condensate before the gravitational growth of perturbations becomes significant~\cite{Lozanov:2019ylm}. Consequently, we assume that the inflaton oscillates in a purely quadratic minimum, or
\begin{equation} 
V(\phi) = \frac{1}{2} m^{2} \phi^{2} 
    \,.
\end{equation}
and that the potential grows more slowly than $\phi^2$ only  at larger values of the field.  The post-inflationary evolution of the homogeneous field is then
\begin{equation} \label{eq:homog_oscil}
\phi = \sqrt{\frac{8}{3}} \frac{M_p}{m} \frac{1}{t} \sin{(m t)}
\end{equation} 
up to arbitrary constants, while $a(t)\sim t^{2/3}$ and the comoving horizon grows as $t^{1/3}$, or $a^{1/2}$. This is equivalent to a matter dominated period of expansion.

For this analysis we need only the initial perturbation spectrum, and we assume that  modes that never left the horizon have vanishing amplitude, while the dimensionless metric perturbations are scale invariant outside the horizon. That is, the power spectrum of the comoving curvature perturbation is  
\begin{equation}
    \label{eq:spectrum}
    \mathcal{P_R}(k)
    =
    \begin{cases}
        A
        & \text{for } k \lesssim k_{\rm horizon}
        \\
        0
        & \text{for } k \gtrsim k_{\rm horizon}
    \end{cases}
    \,.
\end{equation}
The value of  $A$ is sensitive to the form of $V(\phi)$ as inflation ends, which we have not specified, so $A$ is a free parameter.

Perturbations only grow inside the horizon.
Modes just outside the horizon at the end of inflation reenter first and thus  undergo the most growth, with their amplitudes increasing linearly with the expansion of the universe during the linear regime.
We thus expect the first structures to form on comoving scales slightly larger than the Hubble radius at the end of inflation.
Generically, superhorizon modes  will have some scale dependence, but unless this is extreme it will be swamped by extra growth undergone by shorter modes that spend longer inside the horizon.

\section{The Schr\"{o}dinger-Poisson Regime}
\label{sec:scales}

The homogeneous oscillation timescale in equation~\eqref{eq:homog_oscil} is $1/m$.
For our analysis we set $m=\sqrt{3} H_{\rm end}$, where $H_{\rm end}$ is the Hubble parameter at the end of inflation.
The horizon  scales as $1/H \sim a^{3/2}$, while the oscillation time, $1/m$, stays constant.
Consequently, a few efolds after inflation ends, $1/H \gg 1/m$ and  numerically evolving the full Klein-Gordon dynamics for several Hubble times with realistic parameter values is computationally infeasible.
Moreover, at the onset of non-linearity,  perturbations of interest are safely subhorizon, bulk motions are nonrelativistic, and occupation numbers are high.
These are precisely the circumstances in which the (Newtonian) \SP{} formalism is applicable: matter is represented by the non-relativistic wavefunction $\psi$ and the gravitational potential $\Phi$ is found by solving the Poisson equation.

The \SP{} equations are derived by applying the ansatz~\cite{Widrow:1993qq}
\begin{equation}
    \label{eq:phi_transform}
    \phi = \frac{1}{m a^{3/2}} \left( \psi e^{-imt} + \psi^* e^{imt} \right)
\end{equation}
to \cref{eq:KG} and the Einstein field equations, where $\psi$ is a complex variable varying over both time and space. This factors out the homogeneous oscillations in equation~\eqref{eq:homog_oscil}. One then separates out the $e^{\pm imt}$ components  and makes the approximation $m\gg|\dot{\psi}/\psi|$. This resembles the WKB approximation, in that dynamics on the timescale of the condensate oscillations  are assumed  to be ignorable. The result is
\begin{eqnarray}
    \label{eq:schrodinger}
    i \dot{\psi}
    &=&
    - \frac{1}{2ma^2}\nabla^{2} \psi + m \psi \Phi \, ,
    \\ 
    \label{eq:poisson}
    \nabla^{2} \Phi &=& \frac{4 \pi G}{a} \left( |\psi|^2 - \langle |\psi|^2 \rangle \right) \, ,
\end{eqnarray}
and the matter density is  given by $|\psi|^2$. 

The \SP{} equations also describe structure formation with ULDM~\cite{Hu:2000ke, Marsh:2013ywa, Marsh:2015xka, Hui:2016ltb,Coles:2002sj}.
To explore their early universe analogue  we have generalised \textsc{PyUltraLight}~\cite{Edwards:2018ccc} to an expanding background with appropriate initial conditions, along with a dynamical timestep.\footnote{This code will be described in a forthcoming publication.} This code has been run at resolutions of up to $512^3$. The spatial resolution must be sufficient to resolve the change in the phase of $\psi$  at adjacent gridpoints.  One can understand this through the fluid approximation where the velocity is proportional to the gradient of the phase. If the phase changes by more than $\pi$ between any neighbouring grid points, the fluid velocity is not well-defined and the code halts when the phase of any two neighbouring grid points changes by more than $\pi/2$. The simulation volume is a fixed comoving region with periodic boundary conditions.

In the linear regime the density perturbations are
\begin{equation}
    \psi = \sqrt{\rho_c} \sqrt{1 + \delta} e^{iS}
\end{equation}
where $\rho_c$ is the critical density at the end of inflation and the Fourier representations of $\delta$ and $S$ are~\cite{Woo:2008nn}
\begin{gather} 
    \delta_k(a)  =\tilde{A}_kR_k(a) \label{eq:Rk}  \\
    S_k(a)    =  - \frac{d \delta_{k}}{d x} \label{eq:Ik}
\end{gather}
with
\begin{gather}
    R_k(a) =     \left(\frac{3}{x^2}-1\right) \cos(x) + \frac{3}{x} \sin(x) \label{eq:Rk2}
    \\
    x = \frac{k^{2}}{m \sqrt{\Hend^{2} a}}\label{eq:x}
    \,.
\end{gather}
The scale factor is set to unity at the end of inflation.

We set the initial conditions for our simulations by evolving all modes forward from the end of inflation, via \cref{eq:Rk,eq:Ik,eq:Rk2,eq:x}, until shortly before the perturbative description begins to break down.\footnote{Strictly, the \SP{} equations  only become valid a few e-folds after the end of inflation. A full treatment would follow the analysis of Ref.~\cite{Easther:2010mr} through this initial phase but the change in the power spectrum is small and effectively absorbed into the definition of the spectrum and the  parameter $A$.}
The $\tilde{A}_k$ values are set via the power spectrum at the end of inflation. On superhorizon scales they are time independent Fourier components of a random Gaussian field with amplitude $\alpha k^{1/2}/R_k(a=1)$. The constant $\alpha$ is set such that, at $a=1$, $\rms(\delta) \simeq 0.01$ on comoving distance scales of $0.02/H_{\rm end}$.
This large value is chosen for computational convenience in this initial analysis.

\begin{figure}[tb]
    \begin{center}
    \includegraphics[width=0.9\linewidth]{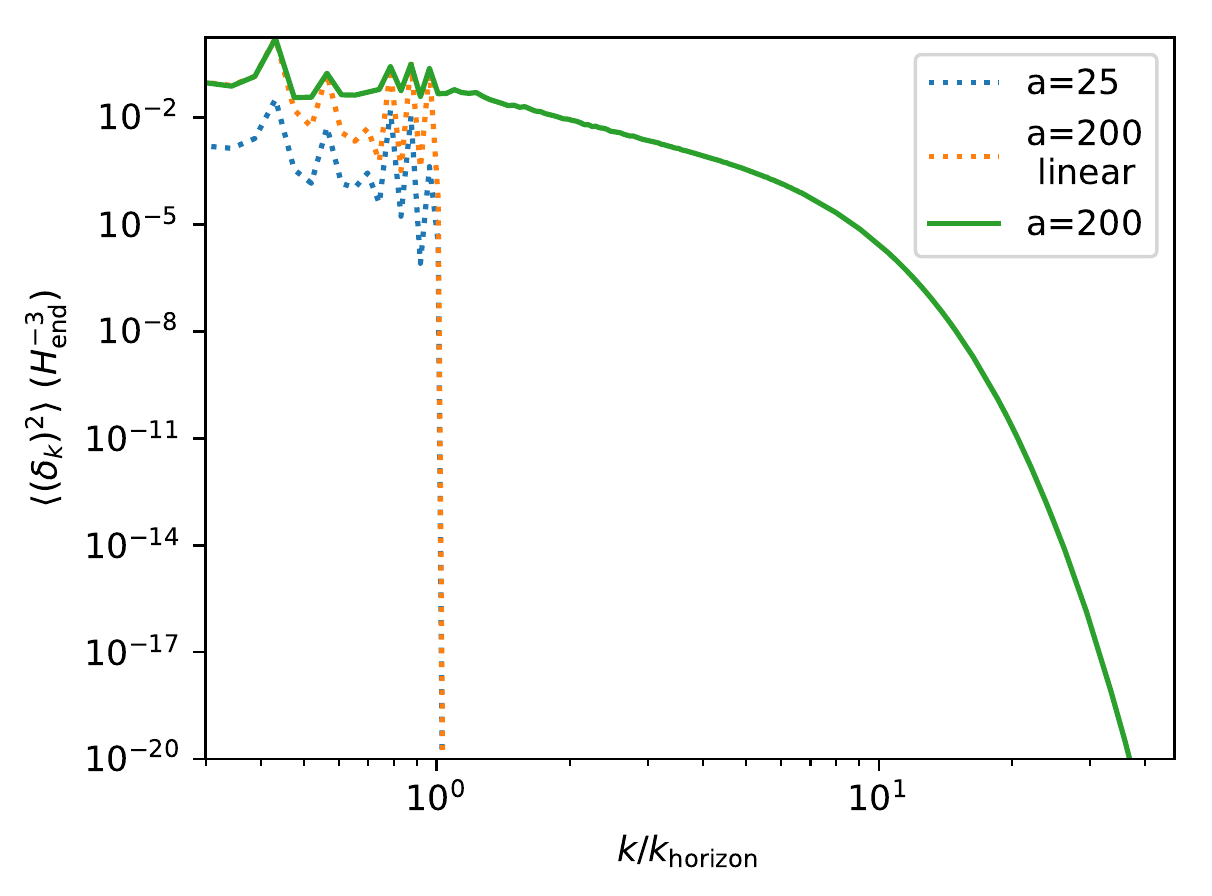}
    \end{center}
    \caption{%
        Power spectrum of the density contrast at the beginning and end of the simulation.
        Scales are given relative to the horizon scale at the end of inflation.  
        }
    \label{fig:spectrum}
\end{figure}

Our ansatz sets the initial subhorizon fluctuation amplitude to zero.
In equation~\eqref{eq:Rk}, when $x\gtrsim1$, $\delta_k$ oscillates, but does not grow.
When $x\lesssim1$, $\delta_k \propto a$.
As inflation ends $x\simeq1$ for $k_{\rm horizon}$; and subhorizon modes are initially oscillatory.
Consequently, it can be shown that unless the initial subhorizon power spectrum grows as fast as $k^8$ it will be sub-dominant when nonlinearity sets in. The initial spectrum and its  evolution is shown in \Cref{fig:spectrum}.

Modes which are mildly superhorizon at the end of inflation become nonlinear first, so our simulation volume must be initially superhorizon.\footnote{This  also occurs in large cosmological N-body simulations; the \SP{} system is a similar Newtonian limit. Likewise, the breakdown of the perturbative description resembles the onset of nonlinearity during structure formation --  the gravitational potential remains small but  density perturbations become large.} Here we take $\Lbox = 10/H_{\rm end}$, in comoving units. For this choice the full box is subhorizon when $a \gtrsim 10^2$, but all structure formation occurs on much smaller scales.

\section{Results}
\label{sec:results}

\begin{figure}[tb]
    \centering
    \includegraphics[width=0.9\linewidth]{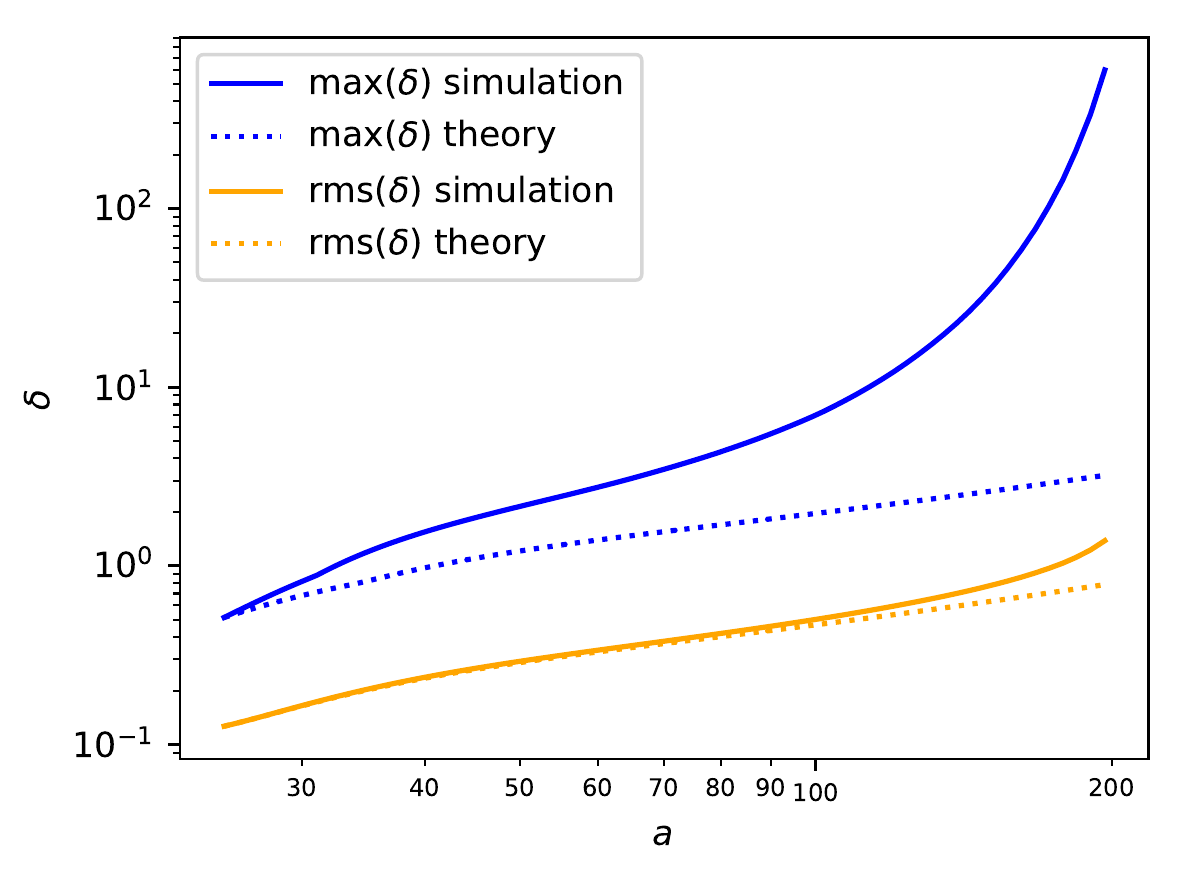}
    \caption{%
        Root mean square and maximum density contrast, measured at comoving distance scale $\simeq 0.02/H_{\rm end}$;  The scale factor $a$ is unity at the end of inflation. The results diverge as $\delta$ increases and perturbation theory breaks down.
    }
    \label{fig:delta_sim_theory}
\end{figure}

We focus on the density contrast $\delta = (\rho - \bar{\rho})/\bar{\rho}$, where $\bar{\rho}$ is the average density.
Previous perturbative treatments of  the inflaton field break down when $\delta \lesssim 1$ but here we numerically evolve the inflaton condensate well into the non-linear regime, reaching a maximum density contrast of $600$ on a $512^3$ grid.
This corresponds to a scale factor of 200, which itself corresponds to $\gtrsim 3000$ oscillations of the homogeneous background solution and since $t \propto a^{3/2}$ most of these oscillations happen near the end of our simulation.
We vary the time step to match the relevant dynamical timescale but a full relativistic simulation would need to resolve all the oscillations.

\Cref{fig:delta_sim_theory} shows both the maximum value and \RMS{} of $\delta$ in the simulation volume as function of the scale factor for a given realisation of our model. We compare the full nonlinear result to the purely perturbative evolution of \cref{eq:Rk} using the same initial configuration.
The root mean square of the density contrast in the simulations clearly departs from the linear result. Moreover, the peak densities correspond to structures with density contrasts of ${\cal{O}}(10^2)$ that have broken away from the ``Hubble flow''.
Note that these results also serve to verify the solver, which we also tested by extracting and plotting individual modes with $\delta\ll1$ and matching them to the perturbation evolution.

\begin{figure}[tb]
    \centering
    \includegraphics[width=0.9\linewidth]{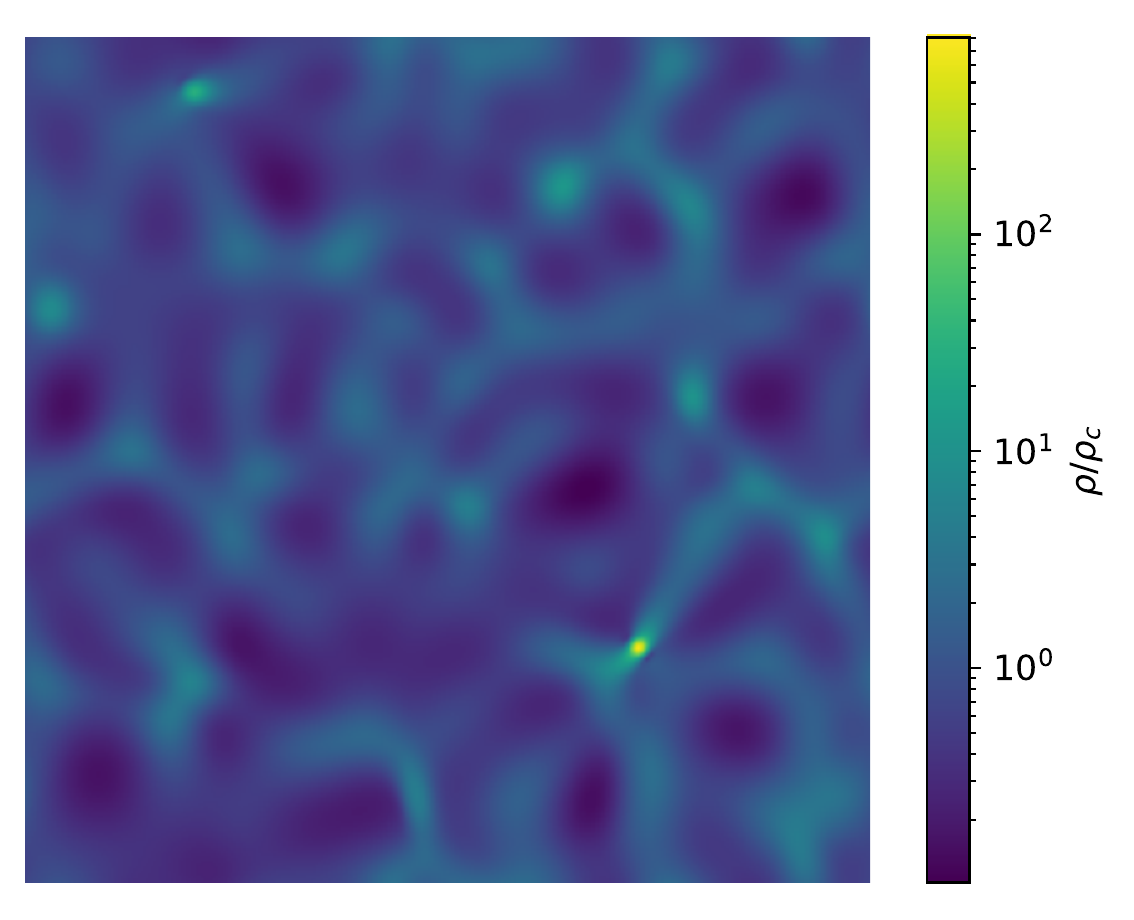} \\
    \includegraphics[width=0.9\linewidth]{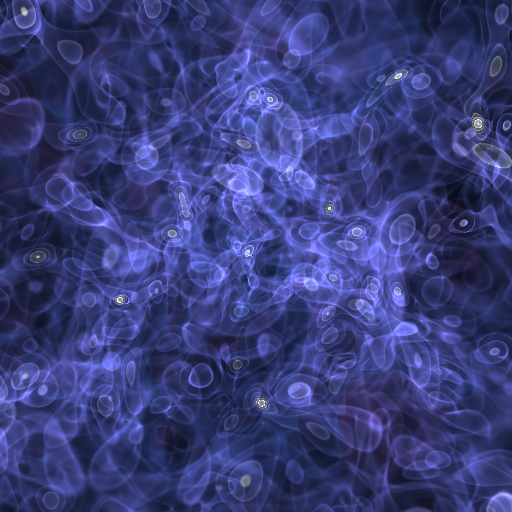}
    \caption{% 
     Simulation results for an initial perturbation of amplitude $\sim10^{-2}$; comoving simulation box size of 10 times post-inflationary  Hubble radius; when the universe has expanded by a factor of $a=200$ since the end of inflation.
        [Top] The density $\rho$ along a slice including the point of highest density.
        [Bottom]  Volume rendering of a subset of the box; blue regions $\delta \sim 1$;  yellow/white regions  $\delta \sim 10-100$.
    }
    \label{fig:volume}
\end{figure}

\Cref{fig:volume} shows a specific configuration, shortly before the phase-gradient condition is violated. Qualitatively, the peaks  are typically aspherical and occur at the intersections of a growing weblike network of overdensities, strongly reminiscent of those seen in standard cosmological structure simulations.
The phase-gradient condition tends to be first violated at large overdensities; it is here that velocities will be largest. These breakdowns  are localised and do not immediately ``propagate'' to the wider grid; running the code past the point at which they  become manifest shows the further development of the web.

Consequently, this analysis confirms expectations that a pure inflaton condensate will fragment gravitationally if no other processes (such as resonance or prompt reheating) disrupt it earlier.
Moreover, it demonstrates that \SP{} solvers can be used to investigate this previously unexplored regime of nonlinear dynamics in the post-inflationary universe.

\section{Discussion}
\label{sec:discussion}

We have made the first exploration of the non-linear gravitational dynamics in the primordial dark age  following inflation in scenarios without resonance~\cite{Amin:2014eta}. We  show  this phase is well-described by the \SP{} equation, solving it  numerically to demonstrate the nonlinear evolution and fragmentation of the inflaton field. 

Beyond the intrinsic interest of nonlinear dynamics in the primordial universe, there are several  mechanisms via which these  can have observable consequences.
For any given inflationary model the ``matching'' between present-day and primordial scales depends on the reheating history, which has a small but potentially detectable impact on the observable perturbation spectrum~\cite{Dodelson:2003vq,Liddle:2003as,Peiris:2008be,Adshead:2010mc} and in curvaton scenarios the duration of the post-inflationary ``matter dominated'' phase is a key parameter~\cite{Lyth:2002my,Byrnes:2014xua}.
If reheating occurs via simple couplings between the inflaton and other species~\cite{Albrecht:1982mp},   particle production scales with the square of the local density and is enhanced by large inhomogeneities.
Moreover, there are many mechanisms via which dark matter can be (over)produced during the primordial dark age.
In some cases, heavy relics  overclose the universe  if the thermalization temperature is  high (e.g.~\cite{Kawasaki:2006gs}); in others  dark matter production directly involves the post-inflationary dynamics~\cite{Chung:1998ua,Liddle:2006qz,Easther:2013nga,Fan:2013faa,Tenkanen:2016jic,Tenkanen:2016twd,Hooper:2018buz,Almeida:2018oid,Tenkanen:2019cik} and could thus be significantly affected by the fragmentation of the condensate.

Collapsing overdensities generically source gravitational waves~\cite{Assadullahi:2009nf,Assadullahi:2009jc,Jedamzik:2010hq} and  nonlinear phases in the early universe can generate stochastic gravitational wave backgrounds~\cite{Khlebnikov:1997di,Easther:2006vd,Amin:2014eta}.
Typical accelerations and the resulting amplitudes produced via gravitational collapse are naively smaller than those from explosive resonance but, more speculatively, this new phase of nonlinear dynamics provides another channel for the production of a primordial gravitational wave background.

We performed simulations for a range of choices for the initial spectrum, and the outcomes did not depend strongly on ansatz used. Higher resolution simulations will be needed to explore the detailed dynamics of the collapsed structures that form after the inflaton condensate fragments, which may include solitons and dynamical oscillon-like structures~\cite{Amin:2010xe,Amin:2010dc,Amin:2011hj,Lozanov:2017hjm,Lozanov:2019ylm}. 

Many lines of enquiry now present themselves.
Obviously, more sophisticated numerical strategies will allow the nonlinear phase to be investigated in greater  detail.
Results for specific inflationary scenarios can be considered, with the initial conditions for the numerical solver  propagated forward from the inflationary phase via perturbation theory~\cite{Jedamzik:2010dq,Easther:2010mr}, along with scenarios where the Compton wavelength is not similar to the comoving horizon size as inflation ends.
Solving the full Einstein equations for the overall evolution is prohibitively expensive and there is no reason to expect models with simple initial perturbation spectra to produce primordial black holes.
However, a more complex scenario suspected of forming singularities could be examined by using the output of a \SP{} simulation to initialise a relativistic solver~\cite{Khlopov:1985jw,Anantua:2008am,Zagorac:2019ekv,Cotner:2018vug,Giblin:2019nuv,Martin:2019nuw}.
Conversely, in models where quanta of the inflation field are produced resonantly the post-resonance universe may be describable using \SP{} dynamics.
Finally, we are considering a class of inflationary models for which ``structure formation'' takes place in the early universe as well as via the canonical post-recombination growth of galactic halos and many of the analytical and numerical tools used to describe the latter are likely to offer insight into the former.

\section*{Acknowledgements}
\label{}

We thank Peter Adshead, Mustafa Amin, Katy Clough, Mateja Gosenca, Emily Kendall, Eugene Lim and Jens Niemeyer for discussions and comments.
The authors acknowledge the Centre for eResearch at the University of Auckland and the Nectar Research Cloud for their help in facilitating this research\footnote{\url{http://www.eresearch.auckland.ac.nz}}.
The authors wish to acknowledge the use of New Zealand eScience Infrastructure (NeSI) high performance computing facilities, consulting support and/or training services as part of this research. New Zealand's national facilities are provided by NeSI and funded jointly by NeSI's collaborator institutions and through the Ministry of Business, Innovation \& Employment's Research Infrastructure programme\footnote{\url{https://www.nesi.org.nz}}.
We  acknowledge support from the Marsden Fund of the Royal Society of New Zealand.
We made use of open source software including Matplotlib, yt, NumPy, SymPy, pyFFTW and NumExpr~\cite{Hunter:2007ouj,Turk:2010ah,vanderWalt:2011bqk,Meurer:2017yhf,gomersall_2016_59508,Frigo:2005zln,cooke_david_2018_1492916}.

\bibliographystyle{apsrev4-2}
\bibliography{autobib.bib}

%apsrev4-2.bst 2019-01-14 (MD) hand-edited version of apsrev4-1.bst
%Control: key (0)
%Control: author (72) initials jnrlst
%Control: editor formatted (1) identically to author
%Control: production of article title (-1) disabled
%Control: page (0) single
%Control: year (1) truncated
%Control: production of eprint (0) enabled
\begin{thebibliography}{81}%
\makeatletter
\providecommand \@ifxundefined [1]{%
 \@ifx{#1\undefined}
}%
\providecommand \@ifnum [1]{%
 \ifnum #1\expandafter \@firstoftwo
 \else \expandafter \@secondoftwo
 \fi
}%
\providecommand \@ifx [1]{%
 \ifx #1\expandafter \@firstoftwo
 \else \expandafter \@secondoftwo
 \fi
}%
\providecommand \natexlab [1]{#1}%
\providecommand \enquote  [1]{``#1''}%
\providecommand \bibnamefont  [1]{#1}%
\providecommand \bibfnamefont [1]{#1}%
\providecommand \citenamefont [1]{#1}%
\providecommand \href@noop [0]{\@secondoftwo}%
\providecommand \href [0]{\begingroup \@sanitize@url \@href}%
\providecommand \@href[1]{\@@startlink{#1}\@@href}%
\providecommand \@@href[1]{\endgroup#1\@@endlink}%
\providecommand \@sanitize@url [0]{\catcode `\\12\catcode `\$12\catcode
  `\&12\catcode `\#12\catcode `\^12\catcode `\_12\catcode `\%12\relax}%
\providecommand \@@startlink[1]{}%
\providecommand \@@endlink[0]{}%
\providecommand \url  [0]{\begingroup\@sanitize@url \@url }%
\providecommand \@url [1]{\endgroup\@href {#1}{\urlprefix }}%
\providecommand \urlprefix  [0]{URL }%
\providecommand \Eprint [0]{\href }%
\providecommand \doibase [0]{https://doi.org/}%
\providecommand \selectlanguage [0]{\@gobble}%
\providecommand \bibinfo  [0]{\@secondoftwo}%
\providecommand \bibfield  [0]{\@secondoftwo}%
\providecommand \translation [1]{[#1]}%
\providecommand \BibitemOpen [0]{}%
\providecommand \bibitemStop [0]{}%
\providecommand \bibitemNoStop [0]{.\EOS\space}%
\providecommand \EOS [0]{\spacefactor3000\relax}%
\providecommand \BibitemShut  [1]{\csname bibitem#1\endcsname}%
\let\auto@bib@innerbib\@empty
%</preamble>
\bibitem [{\citenamefont {Guth}(1981)}]{Guth:1980zm}%
  \BibitemOpen
  \bibfield  {author} {\bibinfo {author} {\bibfnamefont {A.~H.}\ \bibnamefont
  {Guth}},\ }\href {https://doi.org/10.1103/PhysRevD.23.347} {\bibfield
  {journal} {\bibinfo  {journal} {Phys. Rev.}\ }\textbf {\bibinfo {volume}
  {D23}},\ \bibinfo {pages} {347} (\bibinfo {year} {1981})},\ \bibinfo {note}
  {[Adv. Ser. Astrophys. Cosmol.3,139(1987)]}\BibitemShut {NoStop}%
%%CITATION = PHRVA,D23,347;%%
\bibitem [{\citenamefont {Linde}(1982)}]{Linde:1981mu}%
  \BibitemOpen
  \bibfield  {author} {\bibinfo {author} {\bibfnamefont {A.~D.}\ \bibnamefont
  {Linde}},\ }\bibfield  {booktitle} {\emph {\bibinfo {booktitle} {{QUANTUM
  COSMOLOGY}}},\ }\href {https://doi.org/10.1016/0370-2693(82)91219-9}
  {\bibfield  {journal} {\bibinfo  {journal} {Phys. Lett.}\ }\textbf {\bibinfo
  {volume} {108B}},\ \bibinfo {pages} {389} (\bibinfo {year} {1982})},\
  \bibinfo {note} {[Adv. Ser. Astrophys. Cosmol.3,149(1987)]}\BibitemShut
  {NoStop}%
%%CITATION = PHLTA,108B,389;%%
\bibitem [{\citenamefont {Albrecht}\ and\ \citenamefont
  {Steinhardt}(1982)}]{Albrecht:1982wi}%
  \BibitemOpen
  \bibfield  {author} {\bibinfo {author} {\bibfnamefont {A.}~\bibnamefont
  {Albrecht}}\ and\ \bibinfo {author} {\bibfnamefont {P.~J.}\ \bibnamefont
  {Steinhardt}},\ }\href {https://doi.org/10.1103/PhysRevLett.48.1220}
  {\bibfield  {journal} {\bibinfo  {journal} {Phys. Rev. Lett.}\ }\textbf
  {\bibinfo {volume} {48}},\ \bibinfo {pages} {1220} (\bibinfo {year}
  {1982})},\ \bibinfo {note} {[Adv. Ser. Astrophys.
  Cosmol.3,158(1987)]}\BibitemShut {NoStop}%
%%CITATION = PRLTA,48,1220;%%
\bibitem [{\citenamefont {Baumann}(2011)}]{Baumann:2009ds}%
  \BibitemOpen
  \bibfield  {author} {\bibinfo {author} {\bibfnamefont {D.}~\bibnamefont
  {Baumann}},\ }in\ \href {https://doi.org/10.1142/9789814327183_0010} {\emph
  {\bibinfo {booktitle} {{Physics of the large and the small, TASI 09,
  proceedings of the Theoretical Advanced Study Institute in Elementary
  Particle Physics, Boulder, Colorado, USA, 1-26 June 2009}}}}\ (\bibinfo
  {year} {2011})\ pp.\ \bibinfo {pages} {523--686},\ \Eprint
  {https://arxiv.org/abs/0907.5424} {arXiv:0907.5424 [hep-th]} \BibitemShut
  {NoStop}%
%%CITATION = ARXIV:0907.5424;%%
\bibitem [{\citenamefont {Tanabashi}\ \emph {et~al.}(2018)\citenamefont
  {Tanabashi} \emph {et~al.}}]{Tanabashi:2018oca}%
  \BibitemOpen
  \bibfield  {author} {\bibinfo {author} {\bibfnamefont {M.}~\bibnamefont
  {Tanabashi}} \emph {et~al.} (\bibinfo {collaboration} {Particle Data
  Group}),\ }\href {https://doi.org/10.1103/PhysRevD.98.030001} {\bibfield
  {journal} {\bibinfo  {journal} {Phys. Rev.}\ }\textbf {\bibinfo {volume}
  {D98}},\ \bibinfo {pages} {030001} (\bibinfo {year} {2018})}\BibitemShut
  {NoStop}%
%%CITATION = PHRVA,D98,030001;%%
\bibitem [{\citenamefont {Boyle}\ and\ \citenamefont
  {Steinhardt}(2008)}]{Boyle:2005se}%
  \BibitemOpen
  \bibfield  {author} {\bibinfo {author} {\bibfnamefont {L.~A.}\ \bibnamefont
  {Boyle}}\ and\ \bibinfo {author} {\bibfnamefont {P.~J.}\ \bibnamefont
  {Steinhardt}},\ }\href {https://doi.org/10.1103/PhysRevD.77.063504}
  {\bibfield  {journal} {\bibinfo  {journal} {Phys. Rev.}\ }\textbf {\bibinfo
  {volume} {D77}},\ \bibinfo {pages} {063504} (\bibinfo {year} {2008})},\
  \Eprint {https://arxiv.org/abs/astro-ph/0512014} {arXiv:astro-ph/0512014
  [astro-ph]} \BibitemShut {NoStop}%
%%CITATION = ASTRO-PH/0512014;%%
\bibitem [{\citenamefont {Kofman}\ \emph {et~al.}(1994)\citenamefont {Kofman},
  \citenamefont {Linde},\ and\ \citenamefont {Starobinsky}}]{Kofman:1994rk}%
  \BibitemOpen
  \bibfield  {author} {\bibinfo {author} {\bibfnamefont {L.}~\bibnamefont
  {Kofman}}, \bibinfo {author} {\bibfnamefont {A.~D.}\ \bibnamefont {Linde}},\
  and\ \bibinfo {author} {\bibfnamefont {A.~A.}\ \bibnamefont {Starobinsky}},\
  }\href {https://doi.org/10.1103/PhysRevLett.73.3195} {\bibfield  {journal}
  {\bibinfo  {journal} {Phys. Rev. Lett.}\ }\textbf {\bibinfo {volume} {73}},\
  \bibinfo {pages} {3195} (\bibinfo {year} {1994})},\ \Eprint
  {https://arxiv.org/abs/hep-th/9405187} {arXiv:hep-th/9405187 [hep-th]}
  \BibitemShut {NoStop}%
%%CITATION = HEP-TH/9405187;%%
\bibitem [{\citenamefont {Shtanov}\ \emph {et~al.}(1995)\citenamefont
  {Shtanov}, \citenamefont {Traschen},\ and\ \citenamefont
  {Brandenberger}}]{Shtanov:1994ce}%
  \BibitemOpen
  \bibfield  {author} {\bibinfo {author} {\bibfnamefont {Y.}~\bibnamefont
  {Shtanov}}, \bibinfo {author} {\bibfnamefont {J.~H.}\ \bibnamefont
  {Traschen}},\ and\ \bibinfo {author} {\bibfnamefont {R.~H.}\ \bibnamefont
  {Brandenberger}},\ }\href {https://doi.org/10.1103/PhysRevD.51.5438}
  {\bibfield  {journal} {\bibinfo  {journal} {Phys. Rev.}\ }\textbf {\bibinfo
  {volume} {D51}},\ \bibinfo {pages} {5438} (\bibinfo {year} {1995})},\ \Eprint
  {https://arxiv.org/abs/hep-ph/9407247} {arXiv:hep-ph/9407247 [hep-ph]}
  \BibitemShut {NoStop}%
%%CITATION = HEP-PH/9407247;%%
\bibitem [{\citenamefont {Kofman}\ \emph {et~al.}(1997)\citenamefont {Kofman},
  \citenamefont {Linde},\ and\ \citenamefont {Starobinsky}}]{Kofman:1997yn}%
  \BibitemOpen
  \bibfield  {author} {\bibinfo {author} {\bibfnamefont {L.}~\bibnamefont
  {Kofman}}, \bibinfo {author} {\bibfnamefont {A.~D.}\ \bibnamefont {Linde}},\
  and\ \bibinfo {author} {\bibfnamefont {A.~A.}\ \bibnamefont {Starobinsky}},\
  }\href {https://doi.org/10.1103/PhysRevD.56.3258} {\bibfield  {journal}
  {\bibinfo  {journal} {Phys. Rev.}\ }\textbf {\bibinfo {volume} {D56}},\
  \bibinfo {pages} {3258} (\bibinfo {year} {1997})},\ \Eprint
  {https://arxiv.org/abs/hep-ph/9704452} {arXiv:hep-ph/9704452 [hep-ph]}
  \BibitemShut {NoStop}%
%%CITATION = HEP-PH/9704452;%%
\bibitem [{\citenamefont {Lozanov}\ and\ \citenamefont
  {Amin}(2017)}]{Lozanov:2016hid}%
  \BibitemOpen
  \bibfield  {author} {\bibinfo {author} {\bibfnamefont {K.~D.}\ \bibnamefont
  {Lozanov}}\ and\ \bibinfo {author} {\bibfnamefont {M.~A.}\ \bibnamefont
  {Amin}},\ }\href {https://doi.org/10.1103/PhysRevLett.119.061301} {\bibfield
  {journal} {\bibinfo  {journal} {Phys. Rev. Lett.}\ }\textbf {\bibinfo
  {volume} {119}},\ \bibinfo {pages} {061301} (\bibinfo {year} {2017})},\
  \Eprint {https://arxiv.org/abs/1608.01213} {arXiv:1608.01213 [astro-ph.CO]}
  \BibitemShut {NoStop}%
%%CITATION = ARXIV:1608.01213;%%
\bibitem [{\citenamefont {Amin}(2010)}]{Amin:2010xe}%
  \BibitemOpen
  \bibfield  {author} {\bibinfo {author} {\bibfnamefont {M.~A.}\ \bibnamefont
  {Amin}},\ }\href@noop {} {\  (\bibinfo {year} {2010})},\ \Eprint
  {https://arxiv.org/abs/1006.3075} {arXiv:1006.3075 [astro-ph.CO]}
  \BibitemShut {NoStop}%
%%CITATION = ARXIV:1006.3075;%%
\bibitem [{\citenamefont {Amin}\ \emph {et~al.}(2010)\citenamefont {Amin},
  \citenamefont {Easther},\ and\ \citenamefont {Finkel}}]{Amin:2010dc}%
  \BibitemOpen
  \bibfield  {author} {\bibinfo {author} {\bibfnamefont {M.~A.}\ \bibnamefont
  {Amin}}, \bibinfo {author} {\bibfnamefont {R.}~\bibnamefont {Easther}},\ and\
  \bibinfo {author} {\bibfnamefont {H.}~\bibnamefont {Finkel}},\ }\href
  {https://doi.org/10.1088/1475-7516/2010/12/001} {\bibfield  {journal}
  {\bibinfo  {journal} {JCAP}\ }\textbf {\bibinfo {volume} {1012}},\ \bibinfo
  {pages} {001}},\ \Eprint {https://arxiv.org/abs/1009.2505} {arXiv:1009.2505
  [astro-ph.CO]} \BibitemShut {NoStop}%
%%CITATION = ARXIV:1009.2505;%%
\bibitem [{\citenamefont {Amin}\ \emph {et~al.}(2012)\citenamefont {Amin},
  \citenamefont {Easther}, \citenamefont {Finkel}, \citenamefont {Flauger},\
  and\ \citenamefont {Hertzberg}}]{Amin:2011hj}%
  \BibitemOpen
  \bibfield  {author} {\bibinfo {author} {\bibfnamefont {M.~A.}\ \bibnamefont
  {Amin}}, \bibinfo {author} {\bibfnamefont {R.}~\bibnamefont {Easther}},
  \bibinfo {author} {\bibfnamefont {H.}~\bibnamefont {Finkel}}, \bibinfo
  {author} {\bibfnamefont {R.}~\bibnamefont {Flauger}},\ and\ \bibinfo {author}
  {\bibfnamefont {M.~P.}\ \bibnamefont {Hertzberg}},\ }\href
  {https://doi.org/10.1103/PhysRevLett.108.241302} {\bibfield  {journal}
  {\bibinfo  {journal} {Phys. Rev. Lett.}\ }\textbf {\bibinfo {volume} {108}},\
  \bibinfo {pages} {241302} (\bibinfo {year} {2012})},\ \Eprint
  {https://arxiv.org/abs/1106.3335} {arXiv:1106.3335 [astro-ph.CO]}
  \BibitemShut {NoStop}%
%%CITATION = ARXIV:1106.3335;%%
\bibitem [{\citenamefont {Lozanov}\ and\ \citenamefont
  {Amin}(2018)}]{Lozanov:2017hjm}%
  \BibitemOpen
  \bibfield  {author} {\bibinfo {author} {\bibfnamefont {K.~D.}\ \bibnamefont
  {Lozanov}}\ and\ \bibinfo {author} {\bibfnamefont {M.~A.}\ \bibnamefont
  {Amin}},\ }\href {https://doi.org/10.1103/PhysRevD.97.023533} {\bibfield
  {journal} {\bibinfo  {journal} {Phys. Rev.}\ }\textbf {\bibinfo {volume}
  {D97}},\ \bibinfo {pages} {023533} (\bibinfo {year} {2018})},\ \Eprint
  {https://arxiv.org/abs/1710.06851} {arXiv:1710.06851 [astro-ph.CO]}
  \BibitemShut {NoStop}%
%%CITATION = ARXIV:1710.06851;%%
\bibitem [{\citenamefont {Lozanov}\ and\ \citenamefont
  {Amin}(2019)}]{Lozanov:2019ylm}%
  \BibitemOpen
  \bibfield  {author} {\bibinfo {author} {\bibfnamefont {K.~D.}\ \bibnamefont
  {Lozanov}}\ and\ \bibinfo {author} {\bibfnamefont {M.~A.}\ \bibnamefont
  {Amin}},\ }\href@noop {} {\  (\bibinfo {year} {2019})},\ \Eprint
  {https://arxiv.org/abs/1902.06736} {arXiv:1902.06736 [astro-ph.CO]}
  \BibitemShut {NoStop}%
%%CITATION = ARXIV:1902.06736;%%
\bibitem [{\citenamefont {Felder}\ and\ \citenamefont
  {Tkachev}(2008)}]{Felder:2000hq}%
  \BibitemOpen
  \bibfield  {author} {\bibinfo {author} {\bibfnamefont {G.~N.}\ \bibnamefont
  {Felder}}\ and\ \bibinfo {author} {\bibfnamefont {I.}~\bibnamefont
  {Tkachev}},\ }\href {https://doi.org/10.1016/j.cpc.2008.02.009} {\bibfield
  {journal} {\bibinfo  {journal} {Comput. Phys. Commun.}\ }\textbf {\bibinfo
  {volume} {178}},\ \bibinfo {pages} {929} (\bibinfo {year} {2008})},\ \Eprint
  {https://arxiv.org/abs/hep-ph/0011159} {arXiv:hep-ph/0011159 [hep-ph]}
  \BibitemShut {NoStop}%
%%CITATION = HEP-PH/0011159;%%
\bibitem [{\citenamefont {Frolov}(2008)}]{Frolov:2008hy}%
  \BibitemOpen
  \bibfield  {author} {\bibinfo {author} {\bibfnamefont {A.~V.}\ \bibnamefont
  {Frolov}},\ }\href {https://doi.org/10.1088/1475-7516/2008/11/009} {\bibfield
   {journal} {\bibinfo  {journal} {JCAP}\ }\textbf {\bibinfo {volume} {0811}},\
  \bibinfo {pages} {009}},\ \Eprint {https://arxiv.org/abs/0809.4904}
  {arXiv:0809.4904 [hep-ph]} \BibitemShut {NoStop}%
%%CITATION = ARXIV:0809.4904;%%
\bibitem [{\citenamefont {Easther}\ \emph {et~al.}(2010)\citenamefont
  {Easther}, \citenamefont {Finkel},\ and\ \citenamefont
  {Roth}}]{Easther:2010qz}%
  \BibitemOpen
  \bibfield  {author} {\bibinfo {author} {\bibfnamefont {R.}~\bibnamefont
  {Easther}}, \bibinfo {author} {\bibfnamefont {H.}~\bibnamefont {Finkel}},\
  and\ \bibinfo {author} {\bibfnamefont {N.}~\bibnamefont {Roth}},\ }\href
  {https://doi.org/10.1088/1475-7516/2010/10/025} {\bibfield  {journal}
  {\bibinfo  {journal} {JCAP}\ }\textbf {\bibinfo {volume} {1010}},\ \bibinfo
  {pages} {025}},\ \Eprint {https://arxiv.org/abs/1005.1921} {arXiv:1005.1921
  [astro-ph.CO]} \BibitemShut {NoStop}%
%%CITATION = ARXIV:1005.1921;%%
\bibitem [{\citenamefont {Abbott}\ \emph {et~al.}(1982)\citenamefont {Abbott},
  \citenamefont {Farhi},\ and\ \citenamefont {Wise}}]{Abbott:1982hn}%
  \BibitemOpen
  \bibfield  {author} {\bibinfo {author} {\bibfnamefont {L.~F.}\ \bibnamefont
  {Abbott}}, \bibinfo {author} {\bibfnamefont {E.}~\bibnamefont {Farhi}},\ and\
  \bibinfo {author} {\bibfnamefont {M.~B.}\ \bibnamefont {Wise}},\ }\href
  {https://doi.org/10.1016/0370-2693(82)90867-X} {\bibfield  {journal}
  {\bibinfo  {journal} {Phys. Lett.}\ }\textbf {\bibinfo {volume} {117B}},\
  \bibinfo {pages} {29} (\bibinfo {year} {1982})}\BibitemShut {NoStop}%
%%CITATION = PHLTA,117B,29;%%
\bibitem [{\citenamefont {Dolgov}\ and\ \citenamefont
  {Linde}(1982)}]{Dolgov:1982th}%
  \BibitemOpen
  \bibfield  {author} {\bibinfo {author} {\bibfnamefont {A.~D.}\ \bibnamefont
  {Dolgov}}\ and\ \bibinfo {author} {\bibfnamefont {A.~D.}\ \bibnamefont
  {Linde}},\ }\href {https://doi.org/10.1016/0370-2693(82)90292-1} {\bibfield
  {journal} {\bibinfo  {journal} {Phys. Lett.}\ }\textbf {\bibinfo {volume}
  {116B}},\ \bibinfo {pages} {329} (\bibinfo {year} {1982})}\BibitemShut
  {NoStop}%
%%CITATION = PHLTA,116B,329;%%
\bibitem [{\citenamefont {Albrecht}\ \emph {et~al.}(1982)\citenamefont
  {Albrecht}, \citenamefont {Steinhardt}, \citenamefont {Turner},\ and\
  \citenamefont {Wilczek}}]{Albrecht:1982mp}%
  \BibitemOpen
  \bibfield  {author} {\bibinfo {author} {\bibfnamefont {A.}~\bibnamefont
  {Albrecht}}, \bibinfo {author} {\bibfnamefont {P.~J.}\ \bibnamefont
  {Steinhardt}}, \bibinfo {author} {\bibfnamefont {M.~S.}\ \bibnamefont
  {Turner}},\ and\ \bibinfo {author} {\bibfnamefont {F.}~\bibnamefont
  {Wilczek}},\ }\href {https://doi.org/10.1103/PhysRevLett.48.1437} {\bibfield
  {journal} {\bibinfo  {journal} {Phys. Rev. Lett.}\ }\textbf {\bibinfo
  {volume} {48}},\ \bibinfo {pages} {1437} (\bibinfo {year}
  {1982})}\BibitemShut {NoStop}%
%%CITATION = PRLTA,48,1437;%%
\bibitem [{\citenamefont {Turner}(1983)}]{Turner:1983he}%
  \BibitemOpen
  \bibfield  {author} {\bibinfo {author} {\bibfnamefont {M.~S.}\ \bibnamefont
  {Turner}},\ }\href {https://doi.org/10.1103/PhysRevD.28.1243} {\bibfield
  {journal} {\bibinfo  {journal} {Phys. Rev.}\ }\textbf {\bibinfo {volume}
  {D28}},\ \bibinfo {pages} {1243} (\bibinfo {year} {1983})}\BibitemShut
  {NoStop}%
%%CITATION = PHRVA,D28,1243;%%
\bibitem [{\citenamefont {Jedamzik}\ \emph
  {et~al.}(2010{\natexlab{a}})\citenamefont {Jedamzik}, \citenamefont
  {Lemoine},\ and\ \citenamefont {Martin}}]{Jedamzik:2010dq}%
  \BibitemOpen
  \bibfield  {author} {\bibinfo {author} {\bibfnamefont {K.}~\bibnamefont
  {Jedamzik}}, \bibinfo {author} {\bibfnamefont {M.}~\bibnamefont {Lemoine}},\
  and\ \bibinfo {author} {\bibfnamefont {J.}~\bibnamefont {Martin}},\ }\href
  {https://doi.org/10.1088/1475-7516/2010/09/034} {\bibfield  {journal}
  {\bibinfo  {journal} {JCAP}\ }\textbf {\bibinfo {volume} {1009}},\ \bibinfo
  {pages} {034}},\ \Eprint {https://arxiv.org/abs/1002.3039} {arXiv:1002.3039
  [astro-ph.CO]} \BibitemShut {NoStop}%
%%CITATION = ARXIV:1002.3039;%%
\bibitem [{\citenamefont {Easther}\ \emph {et~al.}(2011)\citenamefont
  {Easther}, \citenamefont {Flauger},\ and\ \citenamefont
  {Gilmore}}]{Easther:2010mr}%
  \BibitemOpen
  \bibfield  {author} {\bibinfo {author} {\bibfnamefont {R.}~\bibnamefont
  {Easther}}, \bibinfo {author} {\bibfnamefont {R.}~\bibnamefont {Flauger}},\
  and\ \bibinfo {author} {\bibfnamefont {J.~B.}\ \bibnamefont {Gilmore}},\
  }\href {https://doi.org/10.1088/1475-7516/2011/04/027} {\bibfield  {journal}
  {\bibinfo  {journal} {JCAP}\ }\textbf {\bibinfo {volume} {1104}},\ \bibinfo
  {pages} {027}},\ \Eprint {https://arxiv.org/abs/1003.3011} {arXiv:1003.3011
  [astro-ph.CO]} \BibitemShut {NoStop}%
%%CITATION = ARXIV:1003.3011;%%
\bibitem [{\citenamefont {Clough}\ \emph {et~al.}(2015)\citenamefont {Clough},
  \citenamefont {Figueras}, \citenamefont {Finkel}, \citenamefont {Kunesch},
  \citenamefont {Lim},\ and\ \citenamefont {Tunyasuvunakool}}]{Clough:2015sqa}%
  \BibitemOpen
  \bibfield  {author} {\bibinfo {author} {\bibfnamefont {K.}~\bibnamefont
  {Clough}}, \bibinfo {author} {\bibfnamefont {P.}~\bibnamefont {Figueras}},
  \bibinfo {author} {\bibfnamefont {H.}~\bibnamefont {Finkel}}, \bibinfo
  {author} {\bibfnamefont {M.}~\bibnamefont {Kunesch}}, \bibinfo {author}
  {\bibfnamefont {E.~A.}\ \bibnamefont {Lim}},\ and\ \bibinfo {author}
  {\bibfnamefont {S.}~\bibnamefont {Tunyasuvunakool}},\ }\href
  {https://doi.org/10.1088/0264-9381/32/24/245011} {\bibfield  {journal}
  {\bibinfo  {journal} {Class. Quant. Grav.}\ }\textbf {\bibinfo {volume}
  {32}},\ \bibinfo {pages} {245011} (\bibinfo {year} {2015})},\ \bibinfo {note}
  {[Class. Quant. Grav.32,24(2015)]},\ \Eprint
  {https://arxiv.org/abs/1503.03436} {arXiv:1503.03436 [gr-qc]} \BibitemShut
  {NoStop}%
%%CITATION = ARXIV:1503.03436;%%
\bibitem [{\citenamefont {East}\ \emph {et~al.}(2016)\citenamefont {East},
  \citenamefont {Kleban}, \citenamefont {Linde},\ and\ \citenamefont
  {Senatore}}]{East:2015ggf}%
  \BibitemOpen
  \bibfield  {author} {\bibinfo {author} {\bibfnamefont {W.~E.}\ \bibnamefont
  {East}}, \bibinfo {author} {\bibfnamefont {M.}~\bibnamefont {Kleban}},
  \bibinfo {author} {\bibfnamefont {A.}~\bibnamefont {Linde}},\ and\ \bibinfo
  {author} {\bibfnamefont {L.}~\bibnamefont {Senatore}},\ }\href
  {https://doi.org/10.1088/1475-7516/2016/09/010} {\bibfield  {journal}
  {\bibinfo  {journal} {JCAP}\ }\textbf {\bibinfo {volume} {1609}}\bibfield
  {number} {\bibinfo  {number} { (09)},\ \bibinfo {pages} {010}},\ }\Eprint
  {https://arxiv.org/abs/1511.05143} {arXiv:1511.05143 [hep-th]} \BibitemShut
  {NoStop}%
%%CITATION = ARXIV:1511.05143;%%
\bibitem [{\citenamefont {Clough}(2017)}]{Clough:2017ixw}%
  \BibitemOpen
  \bibfield  {author} {\bibinfo {author} {\bibfnamefont {K.}~\bibnamefont
  {Clough}},\ }\emph {\bibinfo {title} {{Scalar Fields in Numerical General
  Relativity: Inhomogeneous inflation and asymmetric bubble collapse}}},\ \href
  {https://doi.org/10.1007/978-3-319-92672-8} {Ph.D. thesis},\ \bibinfo
  {school} {King's Coll. London}, \bibinfo {address} {Cham} (\bibinfo {year}
  {2017}),\ \Eprint {https://arxiv.org/abs/1704.06811} {arXiv:1704.06811
  [gr-qc]} \BibitemShut {NoStop}%
%%CITATION = ARXIV:1704.06811;%%
\bibitem [{\citenamefont {Ruffini}\ and\ \citenamefont
  {Bonazzola}(1969)}]{Ruffini:1969qy}%
  \BibitemOpen
  \bibfield  {author} {\bibinfo {author} {\bibfnamefont {R.}~\bibnamefont
  {Ruffini}}\ and\ \bibinfo {author} {\bibfnamefont {S.}~\bibnamefont
  {Bonazzola}},\ }\href {https://doi.org/10.1103/PhysRev.187.1767} {\bibfield
  {journal} {\bibinfo  {journal} {Phys. Rev.}\ }\textbf {\bibinfo {volume}
  {187}},\ \bibinfo {pages} {1767} (\bibinfo {year} {1969})}\BibitemShut
  {NoStop}%
%%CITATION = PHRVA,187,1767;%%
\bibitem [{\citenamefont {Spiegel}(1980)}]{Spiegel:1980ykb}%
  \BibitemOpen
  \bibfield  {author} {\bibinfo {author} {\bibfnamefont {E.~A.}\ \bibnamefont
  {Spiegel}},\ }\href {https://doi.org/10.1016/0167-2789(80)90015-9} {\bibfield
   {journal} {\bibinfo  {journal} {Physica}\ }\textbf {\bibinfo {volume}
  {D1}},\ \bibinfo {pages} {236} (\bibinfo {year} {1980})}\BibitemShut
  {NoStop}%
%%CITATION = PHYSA,D1,236;%%
\bibitem [{\citenamefont {Seidel}\ and\ \citenamefont
  {Suen}(1990)}]{Seidel:1990jh}%
  \BibitemOpen
  \bibfield  {author} {\bibinfo {author} {\bibfnamefont {E.}~\bibnamefont
  {Seidel}}\ and\ \bibinfo {author} {\bibfnamefont {W.-M.}\ \bibnamefont
  {Suen}},\ }\href {https://doi.org/10.1103/PhysRevD.42.384} {\bibfield
  {journal} {\bibinfo  {journal} {Phys. Rev.}\ }\textbf {\bibinfo {volume}
  {D42}},\ \bibinfo {pages} {384} (\bibinfo {year} {1990})}\BibitemShut
  {NoStop}%
%%CITATION = PHRVA,D42,384;%%
\bibitem [{\citenamefont {Widrow}\ and\ \citenamefont
  {Kaiser}(1993)}]{Widrow:1993qq}%
  \BibitemOpen
  \bibfield  {author} {\bibinfo {author} {\bibfnamefont {L.~M.}\ \bibnamefont
  {Widrow}}\ and\ \bibinfo {author} {\bibfnamefont {N.}~\bibnamefont
  {Kaiser}},\ }\href@noop {} {\bibfield  {journal} {\bibinfo  {journal}
  {Astrophys. J.}\ }\textbf {\bibinfo {volume} {416}},\ \bibinfo {pages} {L71}
  (\bibinfo {year} {1993})}\BibitemShut {NoStop}%
%%CITATION = ASJOA,416,L71;%%
\bibitem [{\citenamefont {Woo}\ and\ \citenamefont
  {Chiueh}(2009)}]{Woo:2008nn}%
  \BibitemOpen
  \bibfield  {author} {\bibinfo {author} {\bibfnamefont {T.-P.}\ \bibnamefont
  {Woo}}\ and\ \bibinfo {author} {\bibfnamefont {T.}~\bibnamefont {Chiueh}},\
  }\href {https://doi.org/10.1088/0004-637X/697/1/850} {\bibfield  {journal}
  {\bibinfo  {journal} {Astrophys. J.}\ }\textbf {\bibinfo {volume} {697}},\
  \bibinfo {pages} {850} (\bibinfo {year} {2009})},\ \Eprint
  {https://arxiv.org/abs/0806.0232} {arXiv:0806.0232 [astro-ph]} \BibitemShut
  {NoStop}%
%%CITATION = ARXIV:0806.0232;%%
\bibitem [{\citenamefont {Schive}\ \emph {et~al.}(2014)\citenamefont {Schive},
  \citenamefont {Chiueh},\ and\ \citenamefont {Broadhurst}}]{Schive:2014dra}%
  \BibitemOpen
  \bibfield  {author} {\bibinfo {author} {\bibfnamefont {H.-Y.}\ \bibnamefont
  {Schive}}, \bibinfo {author} {\bibfnamefont {T.}~\bibnamefont {Chiueh}},\
  and\ \bibinfo {author} {\bibfnamefont {T.}~\bibnamefont {Broadhurst}},\
  }\href {https://doi.org/10.1038/nphys2996} {\bibfield  {journal} {\bibinfo
  {journal} {Nature Phys.}\ }\textbf {\bibinfo {volume} {10}},\ \bibinfo
  {pages} {496} (\bibinfo {year} {2014})},\ \Eprint
  {https://arxiv.org/abs/1406.6586} {arXiv:1406.6586 [astro-ph.GA]}
  \BibitemShut {NoStop}%
%%CITATION = ARXIV:1406.6586;%%
\bibitem [{\citenamefont {Schwabe}\ \emph {et~al.}(2016)\citenamefont
  {Schwabe}, \citenamefont {Niemeyer},\ and\ \citenamefont
  {Engels}}]{Schwabe:2016rze}%
  \BibitemOpen
  \bibfield  {author} {\bibinfo {author} {\bibfnamefont {B.}~\bibnamefont
  {Schwabe}}, \bibinfo {author} {\bibfnamefont {J.~C.}\ \bibnamefont
  {Niemeyer}},\ and\ \bibinfo {author} {\bibfnamefont {J.~F.}\ \bibnamefont
  {Engels}},\ }\href {https://doi.org/10.1103/PhysRevD.94.043513} {\bibfield
  {journal} {\bibinfo  {journal} {Phys. Rev.}\ }\textbf {\bibinfo {volume}
  {D94}},\ \bibinfo {pages} {043513} (\bibinfo {year} {2016})},\ \Eprint
  {https://arxiv.org/abs/1606.05151} {arXiv:1606.05151 [astro-ph.CO]}
  \BibitemShut {NoStop}%
%%CITATION = ARXIV:1606.05151;%%
\bibitem [{\citenamefont {Mocz}\ \emph {et~al.}(2017)\citenamefont {Mocz},
  \citenamefont {Vogelsberger}, \citenamefont {Robles}, \citenamefont {Zavala},
  \citenamefont {Boylan-Kolchin}, \citenamefont {Fialkov},\ and\ \citenamefont
  {Hernquist}}]{Mocz:2017wlg}%
  \BibitemOpen
  \bibfield  {author} {\bibinfo {author} {\bibfnamefont {P.}~\bibnamefont
  {Mocz}}, \bibinfo {author} {\bibfnamefont {M.}~\bibnamefont {Vogelsberger}},
  \bibinfo {author} {\bibfnamefont {V.~H.}\ \bibnamefont {Robles}}, \bibinfo
  {author} {\bibfnamefont {J.}~\bibnamefont {Zavala}}, \bibinfo {author}
  {\bibfnamefont {M.}~\bibnamefont {Boylan-Kolchin}}, \bibinfo {author}
  {\bibfnamefont {A.}~\bibnamefont {Fialkov}},\ and\ \bibinfo {author}
  {\bibfnamefont {L.}~\bibnamefont {Hernquist}},\ }\href
  {https://doi.org/10.1093/mnras/stx1887} {\bibfield  {journal} {\bibinfo
  {journal} {Mon. Not. Roy. Astron. Soc.}\ }\textbf {\bibinfo {volume} {471}},\
  \bibinfo {pages} {4559} (\bibinfo {year} {2017})},\ \Eprint
  {https://arxiv.org/abs/1705.05845} {arXiv:1705.05845 [astro-ph.CO]}
  \BibitemShut {NoStop}%
%%CITATION = ARXIV:1705.05845;%%
\bibitem [{\citenamefont {Edwards}\ \emph {et~al.}(2018)\citenamefont
  {Edwards}, \citenamefont {Kendall}, \citenamefont {Hotchkiss},\ and\
  \citenamefont {Easther}}]{Edwards:2018ccc}%
  \BibitemOpen
  \bibfield  {author} {\bibinfo {author} {\bibfnamefont {F.}~\bibnamefont
  {Edwards}}, \bibinfo {author} {\bibfnamefont {E.}~\bibnamefont {Kendall}},
  \bibinfo {author} {\bibfnamefont {S.}~\bibnamefont {Hotchkiss}},\ and\
  \bibinfo {author} {\bibfnamefont {R.}~\bibnamefont {Easther}},\ }\href
  {https://doi.org/10.1088/1475-7516/2018/10/027} {\bibfield  {journal}
  {\bibinfo  {journal} {JCAP}\ }\textbf {\bibinfo {volume} {1810}}\bibfield
  {number} {\bibinfo  {number} { (10)},\ \bibinfo {pages} {027}},\ }\Eprint
  {https://arxiv.org/abs/1807.04037} {arXiv:1807.04037 [astro-ph.CO]}
  \BibitemShut {NoStop}%
%%CITATION = ARXIV:1807.04037;%%
\bibitem [{\citenamefont {Li}\ \emph {et~al.}(2018)\citenamefont {Li},
  \citenamefont {Hui},\ and\ \citenamefont {Bryan}}]{Li:2018kyk}%
  \BibitemOpen
  \bibfield  {author} {\bibinfo {author} {\bibfnamefont {X.}~\bibnamefont
  {Li}}, \bibinfo {author} {\bibfnamefont {L.}~\bibnamefont {Hui}},\ and\
  \bibinfo {author} {\bibfnamefont {G.~L.}\ \bibnamefont {Bryan}},\ }\href@noop
  {} {\  (\bibinfo {year} {2018})},\ \Eprint {https://arxiv.org/abs/1810.01915}
  {arXiv:1810.01915 [astro-ph.CO]} \BibitemShut {NoStop}%
%%CITATION = ARXIV:1810.01915;%%
\bibitem [{\citenamefont {Zhang}\ \emph {et~al.}(2018)\citenamefont {Zhang},
  \citenamefont {Liu},\ and\ \citenamefont {Chu}}]{Zhang:2018ghp}%
  \BibitemOpen
  \bibfield  {author} {\bibinfo {author} {\bibfnamefont {J.}~\bibnamefont
  {Zhang}}, \bibinfo {author} {\bibfnamefont {H.}~\bibnamefont {Liu}},\ and\
  \bibinfo {author} {\bibfnamefont {M.-C.}\ \bibnamefont {Chu}},\ }\href@noop
  {} {\  (\bibinfo {year} {2018})},\ \Eprint {https://arxiv.org/abs/1809.09848}
  {arXiv:1809.09848 [astro-ph.CO]} \BibitemShut {NoStop}%
%%CITATION = ARXIV:1809.09848;%%
\bibitem [{\citenamefont {Hu}\ \emph {et~al.}(2000)\citenamefont {Hu},
  \citenamefont {Barkana},\ and\ \citenamefont {Gruzinov}}]{Hu:2000ke}%
  \BibitemOpen
  \bibfield  {author} {\bibinfo {author} {\bibfnamefont {W.}~\bibnamefont
  {Hu}}, \bibinfo {author} {\bibfnamefont {R.}~\bibnamefont {Barkana}},\ and\
  \bibinfo {author} {\bibfnamefont {A.}~\bibnamefont {Gruzinov}},\ }\href
  {https://doi.org/10.1103/PhysRevLett.85.1158} {\bibfield  {journal} {\bibinfo
   {journal} {Phys. Rev. Lett.}\ }\textbf {\bibinfo {volume} {85}},\ \bibinfo
  {pages} {1158} (\bibinfo {year} {2000})},\ \Eprint
  {https://arxiv.org/abs/astro-ph/0003365} {arXiv:astro-ph/0003365 [astro-ph]}
  \BibitemShut {NoStop}%
%%CITATION = ASTRO-PH/0003365;%%
\bibitem [{\citenamefont {Marsh}\ and\ \citenamefont
  {Silk}(2014)}]{Marsh:2013ywa}%
  \BibitemOpen
  \bibfield  {author} {\bibinfo {author} {\bibfnamefont {D.~J.~E.}\
  \bibnamefont {Marsh}}\ and\ \bibinfo {author} {\bibfnamefont
  {J.}~\bibnamefont {Silk}},\ }\href {https://doi.org/10.1093/mnras/stt2079}
  {\bibfield  {journal} {\bibinfo  {journal} {Mon. Not. Roy. Astron. Soc.}\
  }\textbf {\bibinfo {volume} {437}},\ \bibinfo {pages} {2652} (\bibinfo {year}
  {2014})},\ \Eprint {https://arxiv.org/abs/1307.1705} {arXiv:1307.1705
  [astro-ph.CO]} \BibitemShut {NoStop}%
%%CITATION = ARXIV:1307.1705;%%
\bibitem [{\citenamefont {Marsh}(2016)}]{Marsh:2015xka}%
  \BibitemOpen
  \bibfield  {author} {\bibinfo {author} {\bibfnamefont {D.~J.~E.}\
  \bibnamefont {Marsh}},\ }\href
  {https://doi.org/10.1016/j.physrep.2016.06.005} {\bibfield  {journal}
  {\bibinfo  {journal} {Phys. Rept.}\ }\textbf {\bibinfo {volume} {643}},\
  \bibinfo {pages} {1} (\bibinfo {year} {2016})},\ \Eprint
  {https://arxiv.org/abs/1510.07633} {arXiv:1510.07633 [astro-ph.CO]}
  \BibitemShut {NoStop}%
%%CITATION = ARXIV:1510.07633;%%
\bibitem [{\citenamefont {Hui}\ \emph {et~al.}(2017)\citenamefont {Hui},
  \citenamefont {Ostriker}, \citenamefont {Tremaine},\ and\ \citenamefont
  {Witten}}]{Hui:2016ltb}%
  \BibitemOpen
  \bibfield  {author} {\bibinfo {author} {\bibfnamefont {L.}~\bibnamefont
  {Hui}}, \bibinfo {author} {\bibfnamefont {J.~P.}\ \bibnamefont {Ostriker}},
  \bibinfo {author} {\bibfnamefont {S.}~\bibnamefont {Tremaine}},\ and\
  \bibinfo {author} {\bibfnamefont {E.}~\bibnamefont {Witten}},\ }\href
  {https://doi.org/10.1103/PhysRevD.95.043541} {\bibfield  {journal} {\bibinfo
  {journal} {Phys. Rev.}\ }\textbf {\bibinfo {volume} {D95}},\ \bibinfo {pages}
  {043541} (\bibinfo {year} {2017})},\ \Eprint
  {https://arxiv.org/abs/1610.08297} {arXiv:1610.08297 [astro-ph.CO]}
  \BibitemShut {NoStop}%
%%CITATION = ARXIV:1610.08297;%%
\bibitem [{\citenamefont {Amin}\ and\ \citenamefont
  {Mocz}(2019)}]{Amin:2019ums}%
  \BibitemOpen
  \bibfield  {author} {\bibinfo {author} {\bibfnamefont {M.~A.}\ \bibnamefont
  {Amin}}\ and\ \bibinfo {author} {\bibfnamefont {P.}~\bibnamefont {Mocz}},\
  }\href {https://doi.org/10.1103/PhysRevD.100.063507} {\bibfield  {journal}
  {\bibinfo  {journal} {Phys. Rev.}\ }\textbf {\bibinfo {volume} {D100}},\
  \bibinfo {pages} {063507} (\bibinfo {year} {2019})},\ \Eprint
  {https://arxiv.org/abs/1902.07261} {arXiv:1902.07261 [astro-ph.CO]}
  \BibitemShut {NoStop}%
%%CITATION = ARXIV:1902.07261;%%
\bibitem [{\citenamefont {Dodelson}\ and\ \citenamefont
  {Hui}(2003)}]{Dodelson:2003vq}%
  \BibitemOpen
  \bibfield  {author} {\bibinfo {author} {\bibfnamefont {S.}~\bibnamefont
  {Dodelson}}\ and\ \bibinfo {author} {\bibfnamefont {L.}~\bibnamefont {Hui}},\
  }\href {https://doi.org/10.1103/PhysRevLett.91.131301} {\bibfield  {journal}
  {\bibinfo  {journal} {Phys. Rev. Lett.}\ }\textbf {\bibinfo {volume} {91}},\
  \bibinfo {pages} {131301} (\bibinfo {year} {2003})},\ \Eprint
  {https://arxiv.org/abs/astro-ph/0305113} {arXiv:astro-ph/0305113 [astro-ph]}
  \BibitemShut {NoStop}%
%%CITATION = ASTRO-PH/0305113;%%
\bibitem [{\citenamefont {Liddle}\ and\ \citenamefont
  {Leach}(2003)}]{Liddle:2003as}%
  \BibitemOpen
  \bibfield  {author} {\bibinfo {author} {\bibfnamefont {A.~R.}\ \bibnamefont
  {Liddle}}\ and\ \bibinfo {author} {\bibfnamefont {S.~M.}\ \bibnamefont
  {Leach}},\ }\href {https://doi.org/10.1103/PhysRevD.68.103503} {\bibfield
  {journal} {\bibinfo  {journal} {Phys. Rev.}\ }\textbf {\bibinfo {volume}
  {D68}},\ \bibinfo {pages} {103503} (\bibinfo {year} {2003})},\ \Eprint
  {https://arxiv.org/abs/astro-ph/0305263} {arXiv:astro-ph/0305263 [astro-ph]}
  \BibitemShut {NoStop}%
%%CITATION = ASTRO-PH/0305263;%%
\bibitem [{\citenamefont {Adshead}\ \emph {et~al.}(2011)\citenamefont
  {Adshead}, \citenamefont {Easther}, \citenamefont {Pritchard},\ and\
  \citenamefont {Loeb}}]{Adshead:2010mc}%
  \BibitemOpen
  \bibfield  {author} {\bibinfo {author} {\bibfnamefont {P.}~\bibnamefont
  {Adshead}}, \bibinfo {author} {\bibfnamefont {R.}~\bibnamefont {Easther}},
  \bibinfo {author} {\bibfnamefont {J.}~\bibnamefont {Pritchard}},\ and\
  \bibinfo {author} {\bibfnamefont {A.}~\bibnamefont {Loeb}},\ }\href
  {https://doi.org/10.1088/1475-7516/2011/02/021} {\bibfield  {journal}
  {\bibinfo  {journal} {JCAP}\ }\textbf {\bibinfo {volume} {1102}},\ \bibinfo
  {pages} {021}},\ \Eprint {https://arxiv.org/abs/1007.3748} {arXiv:1007.3748
  [astro-ph.CO]} \BibitemShut {NoStop}%
%%CITATION = ARXIV:1007.3748;%%
\bibitem [{\citenamefont {Munoz}\ and\ \citenamefont
  {Kamionkowski}(2015)}]{Munoz:2014eqa}%
  \BibitemOpen
  \bibfield  {author} {\bibinfo {author} {\bibfnamefont {J.~B.}\ \bibnamefont
  {Munoz}}\ and\ \bibinfo {author} {\bibfnamefont {M.}~\bibnamefont
  {Kamionkowski}},\ }\href {https://doi.org/10.1103/PhysRevD.91.043521}
  {\bibfield  {journal} {\bibinfo  {journal} {Phys. Rev.}\ }\textbf {\bibinfo
  {volume} {D91}},\ \bibinfo {pages} {043521} (\bibinfo {year} {2015})},\
  \Eprint {https://arxiv.org/abs/1412.0656} {arXiv:1412.0656 [astro-ph.CO]}
  \BibitemShut {NoStop}%
%%CITATION = ARXIV:1412.0656;%%
\bibitem [{\citenamefont {Chung}\ \emph {et~al.}(1998)\citenamefont {Chung},
  \citenamefont {Kolb},\ and\ \citenamefont {Riotto}}]{Chung:1998ua}%
  \BibitemOpen
  \bibfield  {author} {\bibinfo {author} {\bibfnamefont {D.~J.~H.}\
  \bibnamefont {Chung}}, \bibinfo {author} {\bibfnamefont {E.~W.}\ \bibnamefont
  {Kolb}},\ and\ \bibinfo {author} {\bibfnamefont {A.}~\bibnamefont {Riotto}},\
  }\href {https://doi.org/10.1103/PhysRevLett.81.4048} {\bibfield  {journal}
  {\bibinfo  {journal} {Phys. Rev. Lett.}\ }\textbf {\bibinfo {volume} {81}},\
  \bibinfo {pages} {4048} (\bibinfo {year} {1998})},\ \Eprint
  {https://arxiv.org/abs/hep-ph/9805473} {arXiv:hep-ph/9805473 [hep-ph]}
  \BibitemShut {NoStop}%
%%CITATION = HEP-PH/9805473;%%
\bibitem [{\citenamefont {Easther}\ \emph {et~al.}(2014)\citenamefont
  {Easther}, \citenamefont {Galvez}, \citenamefont {Ozsoy},\ and\ \citenamefont
  {Watson}}]{Easther:2013nga}%
  \BibitemOpen
  \bibfield  {author} {\bibinfo {author} {\bibfnamefont {R.}~\bibnamefont
  {Easther}}, \bibinfo {author} {\bibfnamefont {R.}~\bibnamefont {Galvez}},
  \bibinfo {author} {\bibfnamefont {O.}~\bibnamefont {Ozsoy}},\ and\ \bibinfo
  {author} {\bibfnamefont {S.}~\bibnamefont {Watson}},\ }\href
  {https://doi.org/10.1103/PhysRevD.89.023522} {\bibfield  {journal} {\bibinfo
  {journal} {Phys. Rev.}\ }\textbf {\bibinfo {volume} {D89}},\ \bibinfo {pages}
  {023522} (\bibinfo {year} {2014})},\ \Eprint
  {https://arxiv.org/abs/1307.2453} {arXiv:1307.2453 [hep-ph]} \BibitemShut
  {NoStop}%
%%CITATION = ARXIV:1307.2453;%%
\bibitem [{\citenamefont {Fan}\ and\ \citenamefont
  {Reece}(2013)}]{Fan:2013faa}%
  \BibitemOpen
  \bibfield  {author} {\bibinfo {author} {\bibfnamefont {J.}~\bibnamefont
  {Fan}}\ and\ \bibinfo {author} {\bibfnamefont {M.}~\bibnamefont {Reece}},\
  }\href {https://doi.org/10.1007/JHEP10(2013)124} {\bibfield  {journal}
  {\bibinfo  {journal} {JHEP}\ }\textbf {\bibinfo {volume} {10}},\ \bibinfo
  {pages} {124}},\ \Eprint {https://arxiv.org/abs/1307.4400} {arXiv:1307.4400
  [hep-ph]} \BibitemShut {NoStop}%
%%CITATION = ARXIV:1307.4400;%%
\bibitem [{\citenamefont {Tenkanen}\ and\ \citenamefont
  {Vaskonen}(2016)}]{Tenkanen:2016jic}%
  \BibitemOpen
  \bibfield  {author} {\bibinfo {author} {\bibfnamefont {T.}~\bibnamefont
  {Tenkanen}}\ and\ \bibinfo {author} {\bibfnamefont {V.}~\bibnamefont
  {Vaskonen}},\ }\href {https://doi.org/10.1103/PhysRevD.94.083516} {\bibfield
  {journal} {\bibinfo  {journal} {Phys. Rev.}\ }\textbf {\bibinfo {volume}
  {D94}},\ \bibinfo {pages} {083516} (\bibinfo {year} {2016})},\ \Eprint
  {https://arxiv.org/abs/1606.00192} {arXiv:1606.00192 [astro-ph.CO]}
  \BibitemShut {NoStop}%
%%CITATION = ARXIV:1606.00192;%%
\bibitem [{\citenamefont {Tenkanen}(2019)}]{Tenkanen:2019cik}%
  \BibitemOpen
  \bibfield  {author} {\bibinfo {author} {\bibfnamefont {T.}~\bibnamefont
  {Tenkanen}},\ }\href@noop {} {\  (\bibinfo {year} {2019})},\ \Eprint
  {https://arxiv.org/abs/1905.11737} {arXiv:1905.11737 [astro-ph.CO]}
  \BibitemShut {NoStop}%
%%CITATION = ARXIV:1905.11737;%%
\bibitem [{\citenamefont {Liddle}\ and\ \citenamefont
  {Urena-Lopez}(2006)}]{Liddle:2006qz}%
  \BibitemOpen
  \bibfield  {author} {\bibinfo {author} {\bibfnamefont {A.~R.}\ \bibnamefont
  {Liddle}}\ and\ \bibinfo {author} {\bibfnamefont {L.~A.}\ \bibnamefont
  {Urena-Lopez}},\ }\href {https://doi.org/10.1103/PhysRevLett.97.161301}
  {\bibfield  {journal} {\bibinfo  {journal} {Phys. Rev. Lett.}\ }\textbf
  {\bibinfo {volume} {97}},\ \bibinfo {pages} {161301} (\bibinfo {year}
  {2006})},\ \Eprint {https://arxiv.org/abs/astro-ph/0605205}
  {arXiv:astro-ph/0605205 [astro-ph]} \BibitemShut {NoStop}%
%%CITATION = ASTRO-PH/0605205;%%
\bibitem [{\citenamefont {Tenkanen}(2016)}]{Tenkanen:2016twd}%
  \BibitemOpen
  \bibfield  {author} {\bibinfo {author} {\bibfnamefont {T.}~\bibnamefont
  {Tenkanen}},\ }\href {https://doi.org/10.1007/JHEP09(2016)049} {\bibfield
  {journal} {\bibinfo  {journal} {JHEP}\ }\textbf {\bibinfo {volume} {09}},\
  \bibinfo {pages} {049}},\ \Eprint {https://arxiv.org/abs/1607.01379}
  {arXiv:1607.01379 [hep-ph]} \BibitemShut {NoStop}%
%%CITATION = ARXIV:1607.01379;%%
\bibitem [{\citenamefont {Hooper}\ \emph {et~al.}(2019)\citenamefont {Hooper},
  \citenamefont {Krnjaic}, \citenamefont {Long},\ and\ \citenamefont
  {Mcdermott}}]{Hooper:2018buz}%
  \BibitemOpen
  \bibfield  {author} {\bibinfo {author} {\bibfnamefont {D.}~\bibnamefont
  {Hooper}}, \bibinfo {author} {\bibfnamefont {G.}~\bibnamefont {Krnjaic}},
  \bibinfo {author} {\bibfnamefont {A.~J.}\ \bibnamefont {Long}},\ and\
  \bibinfo {author} {\bibfnamefont {S.~D.}\ \bibnamefont {Mcdermott}},\ }\href
  {https://doi.org/10.1103/PhysRevLett.122.091802} {\bibfield  {journal}
  {\bibinfo  {journal} {Phys. Rev. Lett.}\ }\textbf {\bibinfo {volume} {122}},\
  \bibinfo {pages} {091802} (\bibinfo {year} {2019})},\ \Eprint
  {https://arxiv.org/abs/1807.03308} {arXiv:1807.03308 [hep-ph]} \BibitemShut
  {NoStop}%
%%CITATION = ARXIV:1807.03308;%%
\bibitem [{\citenamefont {Almeida}\ \emph {et~al.}(2019)\citenamefont
  {Almeida}, \citenamefont {Bernal}, \citenamefont {Rubio},\ and\ \citenamefont
  {Tenkanen}}]{Almeida:2018oid}%
  \BibitemOpen
  \bibfield  {author} {\bibinfo {author} {\bibfnamefont {J.~P.~B.}\
  \bibnamefont {Almeida}}, \bibinfo {author} {\bibfnamefont {N.}~\bibnamefont
  {Bernal}}, \bibinfo {author} {\bibfnamefont {J.}~\bibnamefont {Rubio}},\ and\
  \bibinfo {author} {\bibfnamefont {T.}~\bibnamefont {Tenkanen}},\ }\href
  {https://doi.org/10.1088/1475-7516/2019/03/012} {\bibfield  {journal}
  {\bibinfo  {journal} {JCAP}\ }\textbf {\bibinfo {volume} {1903}},\ \bibinfo
  {pages} {012}},\ \Eprint {https://arxiv.org/abs/1811.09640} {arXiv:1811.09640
  [hep-ph]} \BibitemShut {NoStop}%
%%CITATION = ARXIV:1811.09640;%%
\bibitem [{\citenamefont {Akrami}\ \emph {et~al.}(2018)\citenamefont {Akrami}
  \emph {et~al.}}]{Akrami:2018odb}%
  \BibitemOpen
  \bibfield  {author} {\bibinfo {author} {\bibfnamefont {Y.}~\bibnamefont
  {Akrami}} \emph {et~al.} (\bibinfo {collaboration} {Planck}),\ }\href@noop {}
  {\  (\bibinfo {year} {2018})},\ \Eprint {https://arxiv.org/abs/1807.06211}
  {arXiv:1807.06211 [astro-ph.CO]} \BibitemShut {NoStop}%
%%CITATION = ARXIV:1807.06211;%%
\bibitem [{\citenamefont {Coles}\ and\ \citenamefont
  {Spencer}(2003)}]{Coles:2002sj}%
  \BibitemOpen
  \bibfield  {author} {\bibinfo {author} {\bibfnamefont {P.}~\bibnamefont
  {Coles}}\ and\ \bibinfo {author} {\bibfnamefont {K.}~\bibnamefont
  {Spencer}},\ }\href {https://doi.org/10.1046/j.1365-8711.2003.06529.x}
  {\bibfield  {journal} {\bibinfo  {journal} {Mon. Not. Roy. Astron. Soc.}\
  }\textbf {\bibinfo {volume} {342}},\ \bibinfo {pages} {176} (\bibinfo {year}
  {2003})},\ \Eprint {https://arxiv.org/abs/astro-ph/0212433}
  {arXiv:astro-ph/0212433 [astro-ph]} \BibitemShut {NoStop}%
%%CITATION = ASTRO-PH/0212433;%%
\bibitem [{\citenamefont {Amin}\ \emph {et~al.}(2014)\citenamefont {Amin},
  \citenamefont {Hertzberg}, \citenamefont {Kaiser},\ and\ \citenamefont
  {Karouby}}]{Amin:2014eta}%
  \BibitemOpen
  \bibfield  {author} {\bibinfo {author} {\bibfnamefont {M.~A.}\ \bibnamefont
  {Amin}}, \bibinfo {author} {\bibfnamefont {M.~P.}\ \bibnamefont {Hertzberg}},
  \bibinfo {author} {\bibfnamefont {D.~I.}\ \bibnamefont {Kaiser}},\ and\
  \bibinfo {author} {\bibfnamefont {J.}~\bibnamefont {Karouby}},\ }\href
  {https://doi.org/10.1142/S0218271815300037} {\bibfield  {journal} {\bibinfo
  {journal} {Int. J. Mod. Phys.}\ }\textbf {\bibinfo {volume} {D24}},\ \bibinfo
  {pages} {1530003} (\bibinfo {year} {2014})},\ \Eprint
  {https://arxiv.org/abs/1410.3808} {arXiv:1410.3808 [hep-ph]} \BibitemShut
  {NoStop}%
%%CITATION = ARXIV:1410.3808;%%
\bibitem [{\citenamefont {Peiris}\ and\ \citenamefont
  {Easther}(2008)}]{Peiris:2008be}%
  \BibitemOpen
  \bibfield  {author} {\bibinfo {author} {\bibfnamefont {H.}~\bibnamefont
  {Peiris}}\ and\ \bibinfo {author} {\bibfnamefont {R.}~\bibnamefont
  {Easther}},\ }\href {https://doi.org/10.1088/1475-7516/2008/07/024}
  {\bibfield  {journal} {\bibinfo  {journal} {JCAP}\ }\textbf {\bibinfo
  {volume} {0807}},\ \bibinfo {pages} {024}},\ \Eprint
  {https://arxiv.org/abs/0805.2154} {arXiv:0805.2154 [astro-ph]} \BibitemShut
  {NoStop}%
%%CITATION = ARXIV:0805.2154;%%
\bibitem [{\citenamefont {Lyth}\ \emph {et~al.}(2003)\citenamefont {Lyth},
  \citenamefont {Ungarelli},\ and\ \citenamefont {Wands}}]{Lyth:2002my}%
  \BibitemOpen
  \bibfield  {author} {\bibinfo {author} {\bibfnamefont {D.~H.}\ \bibnamefont
  {Lyth}}, \bibinfo {author} {\bibfnamefont {C.}~\bibnamefont {Ungarelli}},\
  and\ \bibinfo {author} {\bibfnamefont {D.}~\bibnamefont {Wands}},\ }\href
  {https://doi.org/10.1103/PhysRevD.67.023503} {\bibfield  {journal} {\bibinfo
  {journal} {Phys. Rev.}\ }\textbf {\bibinfo {volume} {D67}},\ \bibinfo {pages}
  {023503} (\bibinfo {year} {2003})},\ \Eprint
  {https://arxiv.org/abs/astro-ph/0208055} {arXiv:astro-ph/0208055 [astro-ph]}
  \BibitemShut {NoStop}%
%%CITATION = ASTRO-PH/0208055;%%
\bibitem [{\citenamefont {Byrnes}\ \emph {et~al.}(2014)\citenamefont {Byrnes},
  \citenamefont {Cortês},\ and\ \citenamefont {Liddle}}]{Byrnes:2014xua}%
  \BibitemOpen
  \bibfield  {author} {\bibinfo {author} {\bibfnamefont {C.~T.}\ \bibnamefont
  {Byrnes}}, \bibinfo {author} {\bibfnamefont {M.}~\bibnamefont {Cortês}},\
  and\ \bibinfo {author} {\bibfnamefont {A.~R.}\ \bibnamefont {Liddle}},\
  }\href {https://doi.org/10.1103/PhysRevD.90.023523} {\bibfield  {journal}
  {\bibinfo  {journal} {Phys. Rev.}\ }\textbf {\bibinfo {volume} {D90}},\
  \bibinfo {pages} {023523} (\bibinfo {year} {2014})},\ \Eprint
  {https://arxiv.org/abs/1403.4591} {arXiv:1403.4591 [astro-ph.CO]}
  \BibitemShut {NoStop}%
%%CITATION = ARXIV:1403.4591;%%
\bibitem [{\citenamefont {Kawasaki}\ \emph {et~al.}(2006)\citenamefont
  {Kawasaki}, \citenamefont {Takahashi},\ and\ \citenamefont
  {Yanagida}}]{Kawasaki:2006gs}%
  \BibitemOpen
  \bibfield  {author} {\bibinfo {author} {\bibfnamefont {M.}~\bibnamefont
  {Kawasaki}}, \bibinfo {author} {\bibfnamefont {F.}~\bibnamefont
  {Takahashi}},\ and\ \bibinfo {author} {\bibfnamefont {T.~T.}\ \bibnamefont
  {Yanagida}},\ }\href {https://doi.org/10.1016/j.physletb.2006.05.037}
  {\bibfield  {journal} {\bibinfo  {journal} {Phys. Lett.}\ }\textbf {\bibinfo
  {volume} {B638}},\ \bibinfo {pages} {8} (\bibinfo {year} {2006})},\ \Eprint
  {https://arxiv.org/abs/hep-ph/0603265} {arXiv:hep-ph/0603265 [hep-ph]}
  \BibitemShut {NoStop}%
%%CITATION = HEP-PH/0603265;%%
\bibitem [{\citenamefont {Assadullahi}\ and\ \citenamefont
  {Wands}(2009)}]{Assadullahi:2009nf}%
  \BibitemOpen
  \bibfield  {author} {\bibinfo {author} {\bibfnamefont {H.}~\bibnamefont
  {Assadullahi}}\ and\ \bibinfo {author} {\bibfnamefont {D.}~\bibnamefont
  {Wands}},\ }\href {https://doi.org/10.1103/PhysRevD.79.083511} {\bibfield
  {journal} {\bibinfo  {journal} {Phys. Rev.}\ }\textbf {\bibinfo {volume}
  {D79}},\ \bibinfo {pages} {083511} (\bibinfo {year} {2009})},\ \Eprint
  {https://arxiv.org/abs/0901.0989} {arXiv:0901.0989 [astro-ph.CO]}
  \BibitemShut {NoStop}%
%%CITATION = ARXIV:0901.0989;%%
\bibitem [{\citenamefont {Assadullahi}\ and\ \citenamefont
  {Wands}(2010)}]{Assadullahi:2009jc}%
  \BibitemOpen
  \bibfield  {author} {\bibinfo {author} {\bibfnamefont {H.}~\bibnamefont
  {Assadullahi}}\ and\ \bibinfo {author} {\bibfnamefont {D.}~\bibnamefont
  {Wands}},\ }\href {https://doi.org/10.1103/PhysRevD.81.023527} {\bibfield
  {journal} {\bibinfo  {journal} {Phys. Rev.}\ }\textbf {\bibinfo {volume}
  {D81}},\ \bibinfo {pages} {023527} (\bibinfo {year} {2010})},\ \Eprint
  {https://arxiv.org/abs/0907.4073} {arXiv:0907.4073 [astro-ph.CO]}
  \BibitemShut {NoStop}%
%%CITATION = ARXIV:0907.4073;%%
\bibitem [{\citenamefont {Jedamzik}\ \emph
  {et~al.}(2010{\natexlab{b}})\citenamefont {Jedamzik}, \citenamefont
  {Lemoine},\ and\ \citenamefont {Martin}}]{Jedamzik:2010hq}%
  \BibitemOpen
  \bibfield  {author} {\bibinfo {author} {\bibfnamefont {K.}~\bibnamefont
  {Jedamzik}}, \bibinfo {author} {\bibfnamefont {M.}~\bibnamefont {Lemoine}},\
  and\ \bibinfo {author} {\bibfnamefont {J.}~\bibnamefont {Martin}},\ }\href
  {https://doi.org/10.1088/1475-7516/2010/04/021} {\bibfield  {journal}
  {\bibinfo  {journal} {JCAP}\ }\textbf {\bibinfo {volume} {1004}},\ \bibinfo
  {pages} {021}},\ \Eprint {https://arxiv.org/abs/1002.3278} {arXiv:1002.3278
  [astro-ph.CO]} \BibitemShut {NoStop}%
%%CITATION = ARXIV:1002.3278;%%
\bibitem [{\citenamefont {Khlebnikov}\ and\ \citenamefont
  {Tkachev}(1997)}]{Khlebnikov:1997di}%
  \BibitemOpen
  \bibfield  {author} {\bibinfo {author} {\bibfnamefont {S.~Y.}\ \bibnamefont
  {Khlebnikov}}\ and\ \bibinfo {author} {\bibfnamefont {I.~I.}\ \bibnamefont
  {Tkachev}},\ }\href {https://doi.org/10.1103/PhysRevD.56.653} {\bibfield
  {journal} {\bibinfo  {journal} {Phys. Rev.}\ }\textbf {\bibinfo {volume}
  {D56}},\ \bibinfo {pages} {653} (\bibinfo {year} {1997})},\ \Eprint
  {https://arxiv.org/abs/hep-ph/9701423} {arXiv:hep-ph/9701423 [hep-ph]}
  \BibitemShut {NoStop}%
%%CITATION = HEP-PH/9701423;%%
\bibitem [{\citenamefont {Easther}\ \emph {et~al.}(2007)\citenamefont
  {Easther}, \citenamefont {Giblin},\ and\ \citenamefont
  {Lim}}]{Easther:2006vd}%
  \BibitemOpen
  \bibfield  {author} {\bibinfo {author} {\bibfnamefont {R.}~\bibnamefont
  {Easther}}, \bibinfo {author} {\bibfnamefont {J.~T.}\ \bibnamefont {Giblin},
  \bibfnamefont {Jr.}},\ and\ \bibinfo {author} {\bibfnamefont {E.~A.}\
  \bibnamefont {Lim}},\ }\href {https://doi.org/10.1103/PhysRevLett.99.221301}
  {\bibfield  {journal} {\bibinfo  {journal} {Phys. Rev. Lett.}\ }\textbf
  {\bibinfo {volume} {99}},\ \bibinfo {pages} {221301} (\bibinfo {year}
  {2007})},\ \Eprint {https://arxiv.org/abs/astro-ph/0612294}
  {arXiv:astro-ph/0612294 [astro-ph]} \BibitemShut {NoStop}%
%%CITATION = ASTRO-PH/0612294;%%
\bibitem [{\citenamefont {Khlopov}\ \emph {et~al.}(1985)\citenamefont
  {Khlopov}, \citenamefont {Malomed},\ and\ \citenamefont
  {Zeldovich}}]{Khlopov:1985jw}%
  \BibitemOpen
  \bibfield  {author} {\bibinfo {author} {\bibfnamefont {M.}~\bibnamefont
  {Khlopov}}, \bibinfo {author} {\bibfnamefont {B.~A.}\ \bibnamefont
  {Malomed}},\ and\ \bibinfo {author} {\bibfnamefont {I.~B.}\ \bibnamefont
  {Zeldovich}},\ }\href@noop {} {\bibfield  {journal} {\bibinfo  {journal}
  {Mon. Not. Roy. Astron. Soc.}\ }\textbf {\bibinfo {volume} {215}},\ \bibinfo
  {pages} {575} (\bibinfo {year} {1985})}\BibitemShut {NoStop}%
%%CITATION = MNRAA,215,575;%%
\bibitem [{\citenamefont {Anantua}\ \emph {et~al.}(2009)\citenamefont
  {Anantua}, \citenamefont {Easther},\ and\ \citenamefont
  {Giblin}}]{Anantua:2008am}%
  \BibitemOpen
  \bibfield  {author} {\bibinfo {author} {\bibfnamefont {R.}~\bibnamefont
  {Anantua}}, \bibinfo {author} {\bibfnamefont {R.}~\bibnamefont {Easther}},\
  and\ \bibinfo {author} {\bibfnamefont {J.~T.}\ \bibnamefont {Giblin}},\
  }\href {https://doi.org/10.1103/PhysRevLett.103.111303} {\bibfield  {journal}
  {\bibinfo  {journal} {Phys. Rev. Lett.}\ }\textbf {\bibinfo {volume} {103}},\
  \bibinfo {pages} {111303} (\bibinfo {year} {2009})},\ \Eprint
  {https://arxiv.org/abs/0812.0825} {arXiv:0812.0825 [astro-ph]} \BibitemShut
  {NoStop}%
%%CITATION = ARXIV:0812.0825;%%
\bibitem [{\citenamefont {Zagorac}\ \emph {et~al.}(2019)\citenamefont
  {Zagorac}, \citenamefont {Easther},\ and\ \citenamefont
  {Padmanabhan}}]{Zagorac:2019ekv}%
  \BibitemOpen
  \bibfield  {author} {\bibinfo {author} {\bibfnamefont {J.~L.}\ \bibnamefont
  {Zagorac}}, \bibinfo {author} {\bibfnamefont {R.}~\bibnamefont {Easther}},\
  and\ \bibinfo {author} {\bibfnamefont {N.}~\bibnamefont {Padmanabhan}},\
  }\href {https://doi.org/10.1088/1475-7516/2019/06/052} {\bibfield  {journal}
  {\bibinfo  {journal} {JCAP}\ }\textbf {\bibinfo {volume} {1906}}\bibfield
  {number} {\bibinfo  {number} { (06)},\ \bibinfo {pages} {052}},\ }\Eprint
  {https://arxiv.org/abs/1903.05053} {arXiv:1903.05053 [astro-ph.CO]}
  \BibitemShut {NoStop}%
%%CITATION = ARXIV:1903.05053;%%
\bibitem [{\citenamefont {Cotner}\ \emph {et~al.}(2018)\citenamefont {Cotner},
  \citenamefont {Kusenko},\ and\ \citenamefont {Takhistov}}]{Cotner:2018vug}%
  \BibitemOpen
  \bibfield  {author} {\bibinfo {author} {\bibfnamefont {E.}~\bibnamefont
  {Cotner}}, \bibinfo {author} {\bibfnamefont {A.}~\bibnamefont {Kusenko}},\
  and\ \bibinfo {author} {\bibfnamefont {V.}~\bibnamefont {Takhistov}},\ }\href
  {https://doi.org/10.1103/PhysRevD.98.083513} {\bibfield  {journal} {\bibinfo
  {journal} {Phys. Rev.}\ }\textbf {\bibinfo {volume} {D98}},\ \bibinfo {pages}
  {083513} (\bibinfo {year} {2018})},\ \Eprint
  {https://arxiv.org/abs/1801.03321} {arXiv:1801.03321 [astro-ph.CO]}
  \BibitemShut {NoStop}%
%%CITATION = ARXIV:1801.03321;%%
\bibitem [{\citenamefont {Giblin}\ and\ \citenamefont
  {Tishue}(2019)}]{Giblin:2019nuv}%
  \BibitemOpen
  \bibfield  {author} {\bibinfo {author} {\bibfnamefont {J.~T.}\ \bibnamefont
  {Giblin}}\ and\ \bibinfo {author} {\bibfnamefont {A.~J.}\ \bibnamefont
  {Tishue}},\ }\href@noop {} {\  (\bibinfo {year} {2019})},\ \Eprint
  {https://arxiv.org/abs/1907.10601} {arXiv:1907.10601 [gr-qc]} \BibitemShut
  {NoStop}%
%%CITATION = ARXIV:1907.10601;%%
\bibitem [{\citenamefont {Martin}\ \emph {et~al.}(2019)\citenamefont {Martin},
  \citenamefont {Papanikolaou},\ and\ \citenamefont {Vennin}}]{Martin:2019nuw}%
  \BibitemOpen
  \bibfield  {author} {\bibinfo {author} {\bibfnamefont {J.}~\bibnamefont
  {Martin}}, \bibinfo {author} {\bibfnamefont {T.}~\bibnamefont
  {Papanikolaou}},\ and\ \bibinfo {author} {\bibfnamefont {V.}~\bibnamefont
  {Vennin}},\ }\href@noop {} {\  (\bibinfo {year} {2019})},\ \Eprint
  {https://arxiv.org/abs/1907.04236} {arXiv:1907.04236 [astro-ph.CO]}
  \BibitemShut {NoStop}%
%%CITATION = ARXIV:1907.04236;%%
\bibitem [{\citenamefont {Hunter}(2007)}]{Hunter:2007ouj}%
  \BibitemOpen
  \bibfield  {author} {\bibinfo {author} {\bibfnamefont {J.~D.}\ \bibnamefont
  {Hunter}},\ }\href {https://doi.org/10.1109/MCSE.2007.55} {\bibfield
  {journal} {\bibinfo  {journal} {Comput. Sci. Eng.}\ }\textbf {\bibinfo
  {volume} {9}},\ \bibinfo {pages} {90} (\bibinfo {year} {2007})}\BibitemShut
  {NoStop}%
%%CITATION = CSENF,9,90;%%
\bibitem [{\citenamefont {Turk}\ \emph {et~al.}(2011)\citenamefont {Turk},
  \citenamefont {Smith}, \citenamefont {Oishi}, \citenamefont {Skory},
  \citenamefont {Skillman}, \citenamefont {Abel},\ and\ \citenamefont
  {Norman}}]{Turk:2010ah}%
  \BibitemOpen
  \bibfield  {author} {\bibinfo {author} {\bibfnamefont {M.~J.}\ \bibnamefont
  {Turk}}, \bibinfo {author} {\bibfnamefont {B.~D.}\ \bibnamefont {Smith}},
  \bibinfo {author} {\bibfnamefont {J.~S.}\ \bibnamefont {Oishi}}, \bibinfo
  {author} {\bibfnamefont {S.}~\bibnamefont {Skory}}, \bibinfo {author}
  {\bibfnamefont {S.~W.}\ \bibnamefont {Skillman}}, \bibinfo {author}
  {\bibfnamefont {T.}~\bibnamefont {Abel}},\ and\ \bibinfo {author}
  {\bibfnamefont {M.~L.}\ \bibnamefont {Norman}},\ }\href
  {https://doi.org/10.1088/0067-0049/192/1/9} {\bibfield  {journal} {\bibinfo
  {journal} {Astrophys. J. Suppl.}\ }\textbf {\bibinfo {volume} {192}},\
  \bibinfo {pages} {9} (\bibinfo {year} {2011})},\ \Eprint
  {https://arxiv.org/abs/1011.3514} {arXiv:1011.3514 [astro-ph.IM]}
  \BibitemShut {NoStop}%
%%CITATION = ARXIV:1011.3514;%%
\bibitem [{\citenamefont {van~der Walt}\ \emph {et~al.}(2011)\citenamefont
  {van~der Walt}, \citenamefont {Colbert},\ and\ \citenamefont
  {Varoquaux}}]{vanderWalt:2011bqk}%
  \BibitemOpen
  \bibfield  {author} {\bibinfo {author} {\bibfnamefont {S.}~\bibnamefont
  {van~der Walt}}, \bibinfo {author} {\bibfnamefont {S.~C.}\ \bibnamefont
  {Colbert}},\ and\ \bibinfo {author} {\bibfnamefont {G.}~\bibnamefont
  {Varoquaux}},\ }\href {https://doi.org/10.1109/MCSE.2011.37} {\bibfield
  {journal} {\bibinfo  {journal} {Comput. Sci. Eng.}\ }\textbf {\bibinfo
  {volume} {13}},\ \bibinfo {pages} {22} (\bibinfo {year} {2011})},\ \Eprint
  {https://arxiv.org/abs/1102.1523} {arXiv:1102.1523 [cs.MS]} \BibitemShut
  {NoStop}%
%%CITATION = ARXIV:1102.1523;%%
\bibitem [{\citenamefont {Meurer}\ \emph {et~al.}(2017)\citenamefont {Meurer}
  \emph {et~al.}}]{Meurer:2017yhf}%
  \BibitemOpen
  \bibfield  {author} {\bibinfo {author} {\bibfnamefont {A.}~\bibnamefont
  {Meurer}} \emph {et~al.},\ }\href {https://doi.org/10.7717/peerj-cs.103}
  {\bibfield  {journal} {\bibinfo  {journal} {PeerJ Comput. Sci.}\ }\textbf
  {\bibinfo {volume} {3}},\ \bibinfo {pages} {e103} (\bibinfo {year}
  {2017})}\BibitemShut {NoStop}%
%%CITATION = INSPIRE-1620083;%%
\bibitem [{\citenamefont {Gomersall}(2016)}]{gomersall_2016_59508}%
  \BibitemOpen
  \bibfield  {author} {\bibinfo {author} {\bibfnamefont {H.}~\bibnamefont
  {Gomersall}},\ }\href {https://doi.org/10.5281/zenodo.59508} {\bibinfo
  {title} {{pyFFTW}}} (\bibinfo {year} {2016})\BibitemShut {NoStop}%
\bibitem [{\citenamefont {Frigo}\ and\ \citenamefont
  {Johnson}(2005)}]{Frigo:2005zln}%
  \BibitemOpen
  \bibfield  {author} {\bibinfo {author} {\bibfnamefont {M.}~\bibnamefont
  {Frigo}}\ and\ \bibinfo {author} {\bibfnamefont {S.~G.}\ \bibnamefont
  {Johnson}},\ }\href {https://doi.org/10.1109/JPROC.2004.840301} {\bibfield
  {journal} {\bibinfo  {journal} {IEEE Proc.}\ }\textbf {\bibinfo {volume}
  {93}},\ \bibinfo {pages} {216} (\bibinfo {year} {2005})}\BibitemShut
  {NoStop}%
%%CITATION = IEEPA,93,216;%%
\bibitem [{\citenamefont {Cooke}\ \emph {et~al.}(2018)\citenamefont {Cooke},
  \citenamefont {Hochberg}, \citenamefont {Alted}, \citenamefont {Vilata},
  \citenamefont {Wiebe}, \citenamefont {de~Menten}, \citenamefont {Valentino},\
  and\ \citenamefont {McLeod}}]{cooke_david_2018_1492916}%
  \BibitemOpen
  \bibfield  {author} {\bibinfo {author} {\bibfnamefont {D.}~\bibnamefont
  {Cooke}}, \bibinfo {author} {\bibfnamefont {T.}~\bibnamefont {Hochberg}},
  \bibinfo {author} {\bibfnamefont {F.}~\bibnamefont {Alted}}, \bibinfo
  {author} {\bibfnamefont {I.}~\bibnamefont {Vilata}}, \bibinfo {author}
  {\bibfnamefont {M.}~\bibnamefont {Wiebe}}, \bibinfo {author} {\bibfnamefont
  {G.}~\bibnamefont {de~Menten}}, \bibinfo {author} {\bibfnamefont
  {A.}~\bibnamefont {Valentino}},\ and\ \bibinfo {author} {\bibfnamefont
  {R.~A.}\ \bibnamefont {McLeod}},\ }\href
  {https://doi.org/10.5281/zenodo.1492916} {\bibinfo {title} {{NumExpr: Fast
  numerical expression evaluator for NumPy}}} (\bibinfo {year}
  {2018})\BibitemShut {NoStop}%
\end{thebibliography}%

\end{document}